# Critical assessment of the alleged failure of the Classical Nucleation Theory at low temperatures

Daniel R. Cassar, André H. Serra, Oscar Peitl, Edgar D. Zanotto

*Department of Materials Engineering, Federal University of* São *Carlos*

*Graduate Program in Materials Science and Engineering, São Carlos, SP, Brazil*

## Abstract

The Classical Nucleation Theory allegedly fails to describe the temperature dependence of the homogeneous crystal nucleation rates below the temperature of maximum nucleation, $T_{max}$. Possible explanations for this suspected breakdown have been advanced in the literature. However, the simplest hypothesis has never been tested, that it is a byproduct of nucleation datasets that have not reached the steady-state regime. In this work, we tested this possibility by analyzing published nucleation data for oxide supercooled liquids, using only nucleation and viscosity data measured in samples of the same glass batch that also have satisfied a steady-state regime test. Furthermore, all the uncertainty and regression confidence bands were computed and considered. Having this rigorous protocol, among the 6 datasets analyzed, we only found weak evidence supporting the existence of the nucleation break in 2 datasets. Our collective results thus indicate that the break at $T_{max}$ is not a common feature of all glass-formers.

# 1 Introduction

Two inquisitive and relevant observations about crystal nucleation kinetics in glass-forming substances, when analyzed considering the Classical Nucleation Theory (CNT), have been a matter of debate for several decades [1–14].

The first observation is referred to as the nucleation "discrepancy", and is related to a failure of CNT to describe the actual *values* of the homogeneous crystal nucleation rates; a difference of many orders of magnitude [1–6]. This discrepancy happens when some additional considerations are made in using or analyzing experimental data through the lens of CNT, for instance that the effective diffusion coefficient describing the nucleation process is inversely proportional to the shear viscosity and, most importantly, if the interfacial energy between the critical nucleus and the supercooled liquid (σ) is assumed to be size and temperature-independent (capillarity approximation). However, this discrepancy can be eliminated by force fitting a temperature-dependent σ to crystal nucleation data [1,4,6,15].

The second observation is sometimes referred to as the nucleation "break" and is more subtle than the first; it is related to a failure of CNT to describe the *temperature-dependence* of the homogeneous crystal nucleation rates below a certain temperature [4–14], which is reported to happen close to the temperature of maximum crystal nucleation rate, $T_{max}$, and is typically close to the glass transition temperature, $T_g$. We will briefly review explanations given by relevant papers [10–14] on this issue later on in the article in Section 3. This nucleation "break" is the subject of this communication.

In this article, we step back, rethink, and reanalyze two basic elements about crystal nucleation in supercooled liquids that may shed light on the CNT nucleation breakdown problem:

- Nucleation data that have *not reached a steady-state* regime will inevitably introduce errors in any analysis. Some authors [5,16] have argued—but never tested—whether the alleged failure of CNT below $T_{max}$ could simply be due to using nucleation data that have not reached the steady-state condition. The possibility that steady-state crystal nucleation conditions have not been reached in some datasets is because steady-state regime can take a very long time to achieve, especially at temperatures below the glass transition temperature, $T_g$;

- Nucleation data analyzed with diffusivity (viscosity or nucleation time-lag) data measured in samples of *different* glass batches may introduce errors because dynamic processes, such as diffusion, viscous flow, and crystal nucleation rates, are very sensitive to small deviations in composition and impurities, for instance different amounts of residual OH$^-$ [17–20].

With the previous considerations, here we test a simple and yet powerful explanation for the nucleation break: that it results from using nucleation rate data that have not reached steady-state, or from using inadequate combinations of crystal nucleation rate and diffusion data.

# 2 Governing equations

## 2.1 Steady-state nucleation rate

According to the Classical Nucleation Theory [2,21–26], the steady-state rate of homogeneous nucleation, $J_0$, is given by

$$J_0 = \frac{D_J}{d_0^4}\sqrt{\frac{\sigma}{kT}}\exp\left(-\frac{W_c}{kT}\right), \tag{1}$$

where $d_0$ is the size of the diffusing structural units, $D_J$ is the diffusion coefficient controlling the nucleation process, $k$ is the Boltzmann constant, and $T$ is the absolute temperature. The work of formation of a critical nucleus, $W_c$, can be calculated using

$$W_c = \frac{16\pi\sigma^3}{3\Delta G_V^2}. \tag{2}$$

Equation (2) was obtained assuming that the critical nucleus is spherical and isotropic. $\Delta G_V$ is the thermodynamic driving force for crystallization per unit volume of the crystal phase. Therefore, the three major parameters controlling the nucleation kinetics are: *i)* $\Delta G_V$, which can be measured for macroscopic crystals or calculated from thermodynamic parameters [6,27], which bears the implicit assumption that macroscopic thermodynamic properties can be used as a proxy for microscopic thermodynamic properties; *ii)* $D_J$, which could in principle be measured, but is commonly estimated by using easily measured parameters, such as viscosity or nucleation time-lag; and *iii)* σ, which is extremely difficult to measure and is usually assumed to be constant and left as a fitting parameter when analyzing crystal nucleation data. Sometimes σ is inferred using numerical calculations that force theory and experiments to agree, leading to a temperature dependent σ(*T*). Both approaches will be discussed further in this article.

## 2.2 Transient nucleation rate

When crystal nucleation takes place in an isothermal experiment, a certain period is needed before the given system reaches its steady-state regime. During this transient period, the crystal nucleation rate, *J*, is time-dependent and smaller than $J_0$. This period of non-stationary nucleation is usually observed and determined by measuring the change of the number of super-critical nuclei per unit volume with time ($N_V$ versus *t*). $N_V$ is related to *J* by Eq. (3).

$$J(t) = \frac{d}{dt}N_V(t) \tag{3}$$

One common technique used to measure $N_V$ is the double-stage method, also known as the Tammann method. It consists of nucleating crystals at one temperature $T_n$ (usually around the glass transition temperature) and then developing the nuclei at a higher temperature $T_d$. This double treatment is necessary in most cases, when the crystal growth velocity at $T_n$ is not sufficiently high to grow the critical nuclei to sizes that can be detected and measured by microscopy. Detailed information on this procedure and associated errors can be obtained from Ref. [2].

One major issue of the double-stage method is that some supercritical nuclei that were formed during the nucleation treatment at $T_n$ are sub-critical at the development temperature $T_d$. These sub-critical nuclei have a tangible probability of dissolving back to supercooled liquid, which effectively shifts the $N_V$ curve to longer times (although, physically, it is more accurate to say that the crystal nuclei density is depressed). Figure 1 shows a schematic picture of the difference between single- and double-stage treatments, making the distinction between the single-stage induction time ($t_{\text{ind},n}$) and the double-stage induction time ($t_{\text{ind},d}$). Both are obtained from extrapolating back the asymptotic part in the limit of infinite time of the curve until it touches the $x$-axis. The slopes of the linear part linear parts of the $N_V$ versus $t$ plots in Figure 1 give the steady-state nucleation rates, $J_0$. If there are no nucleation-exclusion zones (regions that are depleted in some chemical species) in the material, then the steady-state nucleation rates from single- and double-stage experiments are expected to be the same (for more information, see Chapter 33 in Ref. [28]).

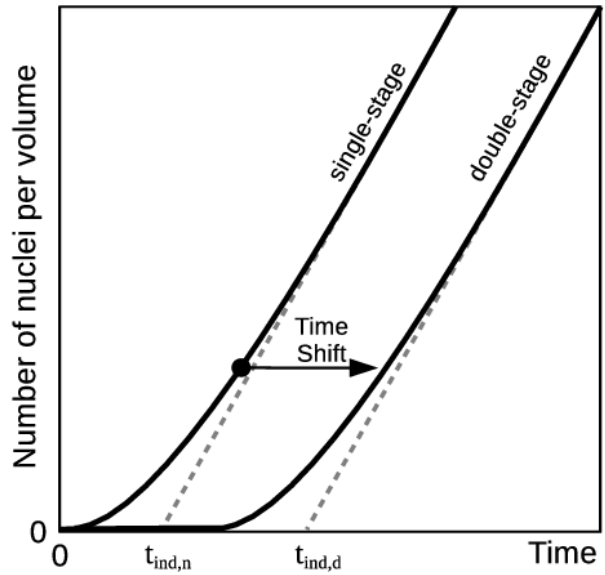

Figure 1    Number of crystals per unit volume versus time at a fixed nucleation temperature for single- and double-stage treatments. The dashed lines are the asymptotic steady-state lines with a slope equal to $J_0$.

Both $t_{\text{ind},n}$ and $t_{\text{ind},d}$ are sensitive to the thermal history of the experiment. In practice, however, few works control and report the thermal history between the glass making procedure and the subsequent crystallization treatments, i.e., the cooling rate of the original melt, heating rate of the glass to the nucleation temperature, $T_n$, and the heating rate to the development temperature, $T_d$. Even the time spent at $T_d$ is sometimes not reported. Because of this problem, noise and distortion in the data are inevitable if variations in the thermal history between samples are significant [29].

Equation (4), proposed by Kashchiev [30], gives a mathematical relation between $N_V$, $J_0$, and the intrinsic nucleation time-lag, $\tau_K$, for a single-stage experiment, where all the supercritical nuclei could be detected, e.g., by a powerful electron microscope. In this case, $\tau_K = 6t_{\text{ind},n}/\pi^2$.

$$N_V = J_0 \tau_K \left[ \frac{t}{\tau_K} - \frac{\pi^2}{6} - 2 \sum_{n=1}^{\infty} \frac{[-1]^n}{n^2} \exp\left(-n^2 \frac{t}{\tau_K}\right) \right] \qquad (4)$$

However, experimentally, it is practically impossible to detect all supercritical nuclei in a single-stage treatment due to the resolution limit of the microscopy technique used to count the nuclei, therefore double-stage treatments are frequently used. Equation (4) may introduce bias in analyzing data collected via double-stage experiments, because these data may have a significant time shift on the $N_V$ curves (Figure 1). In this case, one should use Eq. (5), a modified version of Eq. (4), later proposed by Kashchiev [28], which takes into account the time shift via the parameter $t_g$. In this case, we have $\tau_K = 6(t_{ind,d} - t_g)/\pi^2$.

$$N_V = \begin{cases} 0, t \leq t_g \\ J_0 \tau_K \left[ \frac{t-t_g}{\tau_K} - \frac{\pi^2}{6} - 2 \sum_{n=1}^{\infty} \frac{[-1]^n}{n^2} \exp\left(-n^2 \frac{t-t_g}{\tau_K}\right) \right], t > t_g \end{cases} \quad (5)$$

In the previous equation, $t_g$ is the time necessary for the crystalline clusters to grow from the critical size at the nucleation temperature, $R^*(T_n)$, to the critical size at the development temperature, $R^*(T_d)$. Those nuclei that do not reach $R^*(T_d)$ during the nucleation treatment or during the heating path from $T_n$ to $T_d$, have a significant chance to dissolve back into the supercooled liquid leading to the observed shift in the nucleation plots [29].

In this article, we use Eq. (5) to analyze $N_V$ versus time curves for three glasses ($Li_2Si_2O_5$, $Na_2Ca_2Si_3O_9$, and $Na_4CaSi_3O_9$) because their data were collected using double-stage experiments. We also analyze data collected in single-stage experiments for $Ba_2TiSi_2O_8$, for which we will use Eq. (4).

## 2.3 Considerations using the Classical Nucleation Theory

One unknown parameter in Eq. (1) is the effective diffusion coefficient, $D_J$, which controls the atomic rearrangements involved in crystal nucleation. To compute this parameter, some authors assume that the mechanism that controls $D_J$ also controls the viscous flow, resulting in $D_J \propto D_\eta$, where $D_\eta$ is given by the Stokes–Einstein–Eyring equation (Eq. 6). In other words, this assumption considers that the macro- and micro-rheology are equivalent and are time-independent, (which is likely not true for non-stoichiometric glass-formers, as the composition of the liquid changes when the crystallized fraction changes). Recently, this assumption gained considerable support [31].

$$D_\eta = \phi \frac{kT}{d_0 \eta} \quad (6)$$

In Eq. (6), $\eta$ is the shear viscosity, and $\phi$ is a constant that depends on the assumptions used to derive this equation. If Eyring's approach [32] is used, $\phi = 1$; if the Stokes–Einstein approach [33] is used, $\phi = 1/3\pi$.

Assuming a proportionality constant of unity ($D_J = D_\eta$) combined with Eq. (1) and Eq. (2), we obtain

$$J_0 = \phi \frac{\sqrt{\sigma kT}}{d_0^5 \eta} \exp\left(-\frac{16\pi\sigma^3}{3kT\Delta G_V^2}\right). \quad (7)$$

The jump size parameter, $d_0$, can be estimated by Eq. (8), where $M$ is the molar mass, $\rho$ is the density of the crystal phase, and $N_A$ is Avogadro's number.

$$d_0 = \sqrt[3]{\frac{M}{\rho N_A}} \quad (8)$$

The temperature dependence of the shear viscosity can be obtained from a regression of experimental data using the MYEGA (Mauro–Yue–Ellison–Gupta–Allan) viscosity equation, Eq. (9) [34]. Its three adjustable parameters ($\eta_\infty$, $T_{12}$, and $m$) are defined in Eqs. (10)–(12).

$$\log_{10}(\eta) = \log_{10}(\eta_\infty) + \frac{T_{12}}{T}[12 - \log_{10}(\eta_\infty)] \exp\left(\left[\frac{m}{12-\log_{10}(\eta_\infty)} - 1\right]\left[\frac{T_{12}}{T} - 1\right]\right) \quad (9)$$

$$\eta_\infty \equiv \lim_{T \to \infty} \eta(T) \quad (10)$$

$$\eta(T_{12}) \equiv 10^{12} \text{ Pa.s} \quad (11)$$

$$m \equiv \left.\frac{\partial \log_{10}(\eta)}{\partial \frac{T_{12}}{T}}\right|_{T=T_{12}} \quad (12)$$

With the above equations, the only unknown parameter of Eq. (7) is the interfacial energy, σ. If we assume that σ is *not* temperature dependent, Eq. (7) can be linearized to give:

$$\underbrace{\ln\left(\frac{J_0 \eta}{\sqrt{T}}\right)}_{y} = \underbrace{\ln\left(\frac{\phi\sqrt{\sigma k}}{d_0^5}\right)}_{C_1} - \underbrace{\frac{16\pi\sigma^3}{3k}}_{C_2} \underbrace{\left[\frac{1}{T\Delta G_V^2}\right]}_{x} \quad (13)$$

However, σ is expected to have a weak monotonic temperature dependence [16,35]. This is because it depends on the curvature of the nucleus–liquid interface, thus depending on the critical nucleus radius, which, in turn, varies with temperature [36]. Unfortunately, it is not possible to solve Eq. (7) analytically with respect to σ. Therefore, we used a numerical approach to solve it via the Lambert *W* function, Eq. (14) [15]. The Lambert *W* function is defined as the inverse function of $f(z) = z \exp(z)$, with z being any complex number that yields $z = W(z \exp(z))$. For a more detailed description of Lambert *W* applied to solve CNT with respect of σ, the reader is referred to [15].

$$\sigma = \sqrt[3]{-\frac{kT\Delta G_V^2}{32\pi} W_{-1}\left(-\frac{32\pi}{\Delta G_V^2}\left[\frac{1}{kT}\right]^4 \left[\frac{J_0 \eta d_0^5}{\phi}\right]^6\right)} \quad (14)$$

By assuming a temperature-independent σ to linearize the CNT equation (Eq. 13), or by using the experimental nucleation data to compute the temperature dependence of σ (Eq. 14), we are assuming that the Classical Nucleation Theory is valid. In doing this, we can no longer discuss if CNT is valid or not, but only if it is consistent with experimental data. In other words, a deviation of linearity in Eq. (13) or a non-monotonic temperature dependency of σ when using Eq. (14), can only assess if the considered framework is self-consistent, and not if CNT is valid. New insights on the physics of nucleation, or computational modeling of σ are required for a proper test of CNT. It is important to highlight that this is not a particular limitation of this work, but a general issue in the field of crystal nucleation.

## 3 Motivation and objective

Figure 2a and Figure 2b show plots constructed using literature data for $Li_2Si_2O_5$ [37,38], with Eqs. (13) and (14), respectively. Both plots were produced in this work using literature values of $J_0$ [37,38] and viscosity [5,17,39,40] (shown in Figure 2c).

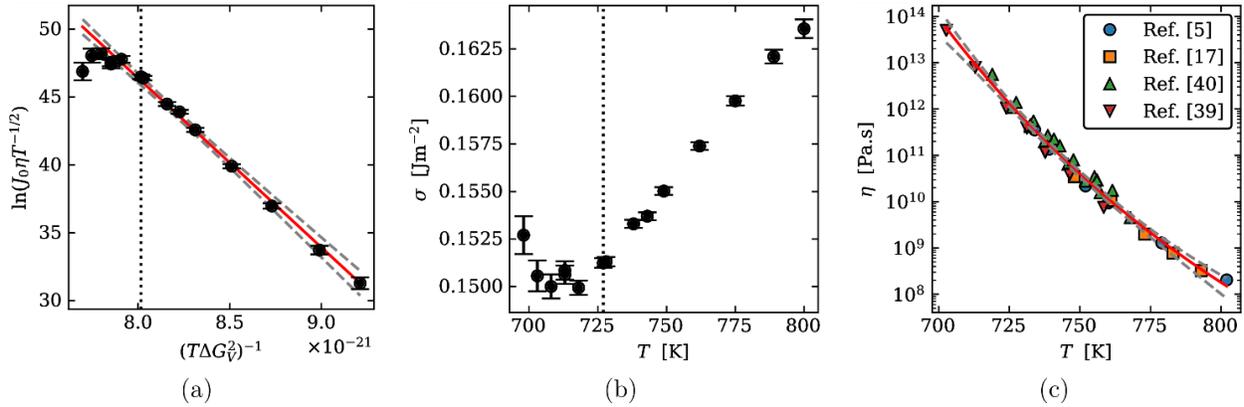

Figure 2   Analysis of the nucleation rate data reported by James [38] for $Li_2Si_2O_5$ **(a)** considering that σ is temperature-independent, and **(b)** the calculated value of σ(T) to force agreement with the experimental nucleation rates. The vertical dotted line marks $T_{max}$ for this dataset. The red line in (a) is the linear regression and the confidence bands are given by the dashed gray line (for more information on the computation of the confidence bands see Section 4.4). **(c)** Viscosity data from various authors [5,17,39,40]; the MYEGA regression is shown by a red line, in which the dashed gray line is the confidence band.

It is important to note that in Figure 2c the viscosity was not measured for samples of the same glass batch as the crystal nucleation rates. The deviation from linearity observed on the low temperature side in Figure 2a, and the non-monotonic nature of the interfacial energy in Figure 2b are manifestations of the nucleation "break". Similar results have been reported for other oxide glasses [4–14]. Most of the reported evidence shows the "break" based on mixed (published) values of $J_0$, $t_{ind,d}$, or η data measured using samples of *different* glass batches.

Possible explanations for the "break" have been advanced by different authors [10–14]. They discuss the thermodynamics and possible formation of metastable phases [10]; errors in measurement of nucleation rates [10]; low temperature relaxation kinetics [10]; the possible role of dynamic heterogeneities in crystal nucleation kinetics [13]; the effect of elastic stresses on the thermodynamic barrier for crystal nucleation [11]; the variation of the size of the "structural units" with temperature [12]; and an effect of the heterogeneous structure of glass-forming liquids that have rigid and floppy regions [14]. Among all these possibilities, it was clearly demonstrated that elastic strains cannot explain the reported break [11], and it was argued that metastable phases, errors in measurements, and low temperature relaxation are likely not the cause [10]; but all the other possibilities are quite reasonable.

Despite that several researchers have observed a break in the experimental nucleation rates at $T_{max}$ when compared to theoretical calculations via the CNT, and some detailed studies have been carried out in a quest to find an explanation for that break, the simplest of all (that the steady-state regime has not been reached in most experiments below $T_{max}$) has not been tested. Also, the use of "mixed" datasets, i.e. nucleation rates from one study (one glass batch) with viscosity data from another work (using samples of a different glass batch of same nominal composition) further complicates this issue.

Therefore, the objective of this work is to test whether the nucleation "break" persists after a more rigorous evaluation of available crystal nucleation rate and viscosity data. We aim to test a simple hypothesis: that the reported break is a byproduct of certain nucleation datasets that have *not* reached the steady-state regime. This evaluation method comprises:

i. Collecting $N_V$ versus time and viscosity data measured for samples from the same glass batch, which we call "clean" datasets. We stress that *all* datasets considered and shown here (except those of Fig. 2) are "clean" by this definition;

ii. Testing and labeling all $N_V$ datasets for those which are probable to have reached the steady-state regime and those that are probable to have not reached the steady-state regime. Those datasets that are labeled as not having reached the steady-state regime are shown to the reader, but never used in the calculations. No data is hidden from the reader in the interest of giving a holistic view of the analysis;

iii. Performing regressions of the $N_V$ versus time datasets accounting for the uncertainty in the adjustable parameters and their confidence bands;

iv. Testing whether the alleged nucleation "break" endures after the previous steps, using Eqs. (13) and (14) and the concepts discussed previously.

# 4 Materials and methods

## 4.1 Materials

The materials of choice for this work were the glass-formers $Li_2Si_2O_5$, $Na_4CaSi_3O_9$, $Na_2Ca_2Si_3O_9$, and $Ba_2TiSi_2O_8$. These are well-documented stoichiometric compositions that undergo homogeneous crystal nucleation when properly heated, and for which enough thermodynamic and kinetic data are available [5,41–51]. $Li_2Si_2O_5$ is considered a model glass for crystallization studies [42], and several works [10–14] reported the nucleation "break" for this material. Weinberg and Zanotto [10] described the break for $Na_4CaSi_3O_9$ and $Na_2Ca_2Si_3O_9$.

The shear viscosity for the $Li_2Si_2O_5$ glass of Serra et al. [50] was determined in the range of $10^{11.5}$ to $10^{14}$ Pa.s using a homemade penetration viscosimeter with a rigid Nimonic indenter. The measurement variables (pressure and time) are similar to that described by Sipp et al. [52]. The experiment to obtain the highest viscosity lasted for 24 hours using an external pressure of 70 MPa.

## 4.2 Literature data collection and grouping strategy

In this work, we thoroughly revisited published crystal nucleation data [39,40,47,50,51]. More specifically, we collected the original $N_V$ versus time data and analyzed them by performing regressions with the modified Kashchiev (Eq. 5) expression. This equation was chosen because most data studied in this work were measured using the double-stage treatment technique. For $Ba_2TiSi_2O_8$, we used Eq. (4) because its dataset was obtained from single-stage treatments.

We should emphasize a key concept used in this article, which we call a "clean" dataset. The adopted strategy was to keep separate (in the analysis) the datasets from different glass batches.

The reasoning behind this choice is that each glass is *unique* because it contains different amounts and types of impurities (including OH⁻ content), as well as some deviation from the respective nominal composition. Some properties are more or less affected by these deviations. For example, density, heat capacity, thermal expansion coefficient, Young`s modulus, and the thermodynamic driving force for crystallization are not significantly affected, whereas dynamic properties, such as viscosity and crystal nucleation kinetics can be strongly affected by small compositional deviations and impurities.

## 4.3 Steady-state regime test

A key question regarding crystal nucleation studies is whether the steady-state regime has been reached for any given $N_V$ dataset, which is especially relevant for studies below the glass transition temperature, because the nucleation time-lags can be quite long. A $N_V$ dataset is defined in this work as a collection of $N_V$ versus time data measured for samples of the same glass batch with the same $T_d$ and $T_n$.

To the best of our knowledge, there is no steady-state test available in the literature. Shneidman made some considerations [53], suggesting that experimental data for which the ratio of $J/J_0$ is greater than 0.93 are good enough to determine the asymptotic form of $N_V$ with "sufficient reliability". While this consideration was not done for the modified Kashchiev equation, we can follow the same reasoning to derive the inequality shown in (15). Figure 3 illustrates the Kashchiev master-curve with the criterion set by the inequality (15).

$$\frac{t-t_g}{\tau_K} > 3.3 \qquad (15)$$

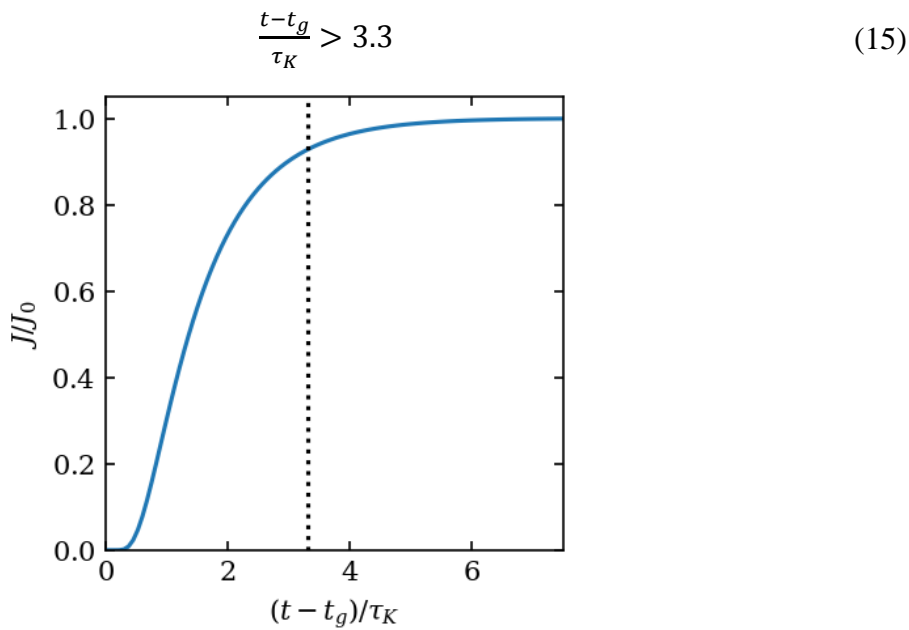

Figure 3  Normalized values of $J/J_0$ for the modified Kashchiev equation. The dotted vertical line shows the position of the reduced critical time of the steady-state test proposed here, based on Shneidman`s work.

One strategy used in this work was to identify the $N_V$ versus time datasets (for each glass) that have a significant chance of <u>not</u> having reached the steady-state regime; i.e., those that do not have a single data point that satisfies the inequality shown in (15). We are aware that the datasets that pass the proposed condition are not *guaranteed* to have reached the steady-state condition. However, if there is not even one experimental point that satisfies inequality (15), the chances are that the steady-state regime has not been achieved.

We will show and discuss all data in the next sections, but some analyses will only be carried out for the datasets that passed the steady-state test.

## 4.4 Numerical calculations

We performed non-linear regressions of Eq. (5) for all the available $N_V$ datasets, except for $Ba_2TiSi_2O_8$, for which we used Eq. (4), as justified in Section 2.2. To improve the reproducibility of this research, the $N_V$ regressions were done following the procedure available in the free and open source module GlassPy [54]. The parameters $J_0$, $\tau_K$, and $t_g$ were considered as free adjustable parameters. After the regression, if the value of $t_g$ was less than 1 second or its standard deviation was greater than the value of $t_g$ itself, then it was fixed to zero and a new regression was made. After completing the new regression, if the value of $\tau_K$ was less than 1 second, then it was also fixed to zero and another regression was performed. The reasoning for this procedure of fixing some parameters to zero is that there was not enough precision to allocate a positive number to these parameters in these cases. This issue can be minimized by having $N_V$ datasets with more experimental data points.

After obtaining the steady-state nucleation rates and respective induction times from the fitting procedure, we applied the steady-state test discussed in Section 4.3. $Li_2Si_2O_5$ data from Fokin, Sycheva, and Serra [39,40,50] were analyzed separately, according to the "clean" analysis concept. We carried out two tests under the following conditions:

i.  Assuming $D_J = D_\eta$ and a temperature-independent $\sigma$. A linear correlation between $y$ and $x$ is expected if CNT and data are consistent with these assumptions (see Eq. 13);

ii. Assuming $D_J = D_\eta$ and a temperature-dependent $\sigma$. A monotonic temperature dependence [16,35] of $\sigma$ is expected if CNT and data are consistent with these assumptions (see Eq. 14).

We are not arguing which of the previous considerations is the best evidence of the possible nucleation "break". Instead, we propose to check both to build a more comprehensive picture of this problem.

Uncertainty propagation was carried out by the linear error propagation theory, computed using the Python uncertainties module [55]. The regression confidence bands were computed using LMFIT`s implementation [56] of the procedure described in Ref. [57], which, in turn, refers to the work of Wolberg [58]. A confidence level of 95% was used in all statistical calculations in this work, except if stated otherwise.

# 5 Results

## 5.1 Viscosity and crystal nucleation rates

Figure 4 shows the viscosity data, the resulting regressions of Eq. (9), and the respective confidence bands. The reasoning for performing the regressions of $Li_2Si_2O_5$ separately for each author is to follow the "clean" dataset approach defined in Section 4.2.

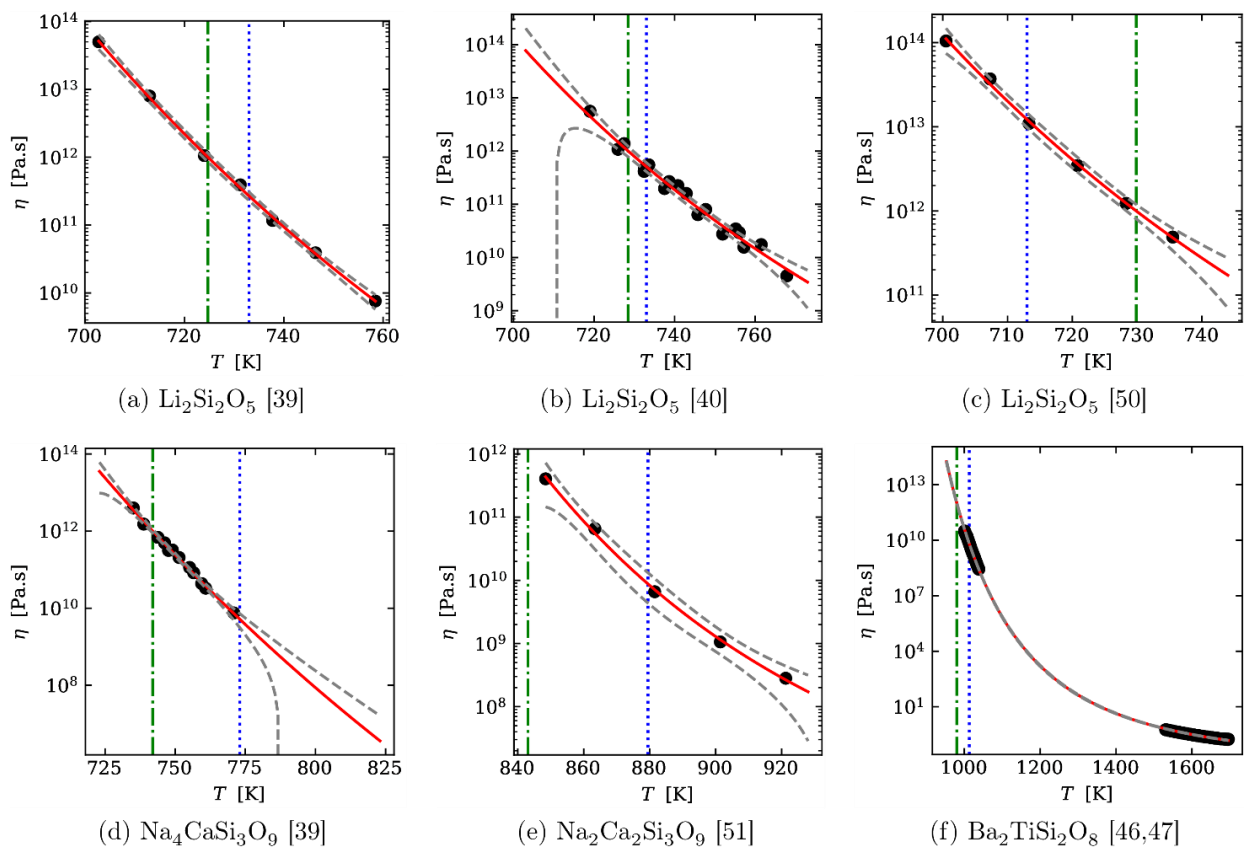

Figure 4   Temperature dependence of the shear viscosity. The black circles are experimental data, the red continuous lines are the regressions of Eq. (9), and the dashed grey lines are the regression confidence bands. The dotted vertical lines are the nucleation maxima, $T_{max}$ (see Figure 5), and the dash-dotted vertical lines refer to $T_{12}$. The temperature range of each plot encompasses the range of nucleation rate data available for each glass.

Table 1 shows the resulting parameters and respective standard deviations obtained from the regressions of viscosity data of Figure 4. As expected, the uncertainty in $\eta_\infty$ is substantial when high temperature viscosity data near the melting point are missing, which happens for all these compositions except $Ba_2TiSi_2O_8$. This uncertainty can be problematic if the regressions are extrapolated too far from the temperature domain of available viscosity data. To visualize this problem, some subplots in Figure 4 show extrapolations of the viscosity regression to cover the temperature domain of experimental nucleation data. Other subplots do not show any extrapolation of the viscosity regression because the temperature domain of nucleation data is contained within the temperature domain of viscosity data. For instance, the poor confidence of extrapolation of data in Figure 4d for high temperatures will result in a larger uncertainty in the analyses that rely on this particular extrapolation (see the right hand side of Figures 5d and 6d).

Table 1. Viscosity parameters ($T_{12}$, $m$, and $\log_{10}(\eta_\infty)$) obtained from regressions using Eq. (9). The uncertainty is one standard deviation. The temperature $T_\Omega$ is defined in Section 6. Units in SI.

| Composition | $T_{12}$ | $m$ | $\log_{10}(\eta_\infty)$ | $T_\Omega$ |
|---|---|---|---|---|
| Li$_2$Si$_2$O$_5$ [39] | 724.7(3) | 51.6(7) | −1(3) | 224 |
| Li$_2$Si$_2$O$_5$ [40] | 728.5(7) | 48(4) | 1(9) | 277 |
| Li$_2$Si$_2$O$_5$ [50] | 729.9(4) | 43(3) | 5(3) | 302 |
| Na$_4$CaSi$_3$O$_9$ [39] | 742.0(3) | 58(2) | −20(40) | 231 |
| Na$_2$Ca$_2$Si$_3$O$_9$ [51] | 843(1) | 59(5) | 5(1) | 332 |
| Ba$_2$TiSi$_2$O$_8$ [46,47] | 980.2(2) | 73.1(4) | −2.27(3) | 586 |

Figure 5 shows the values of steady-state nucleation rates, $J_0$. From these plots, we defined $T_{max}$ (indicated as a vertical dotted line) as the temperature with the highest value of $J_0$, without considering the uncertainty. All plots showing the regressions of $N_V$ versus $t$ data are detailed in the Supplementary Material. The gray circles in Figure 5 refer to $N_V$ datasets that have reached the steady-state (see Section 4.3), whereas the orange squares are those that have not reached the steady-state regime.

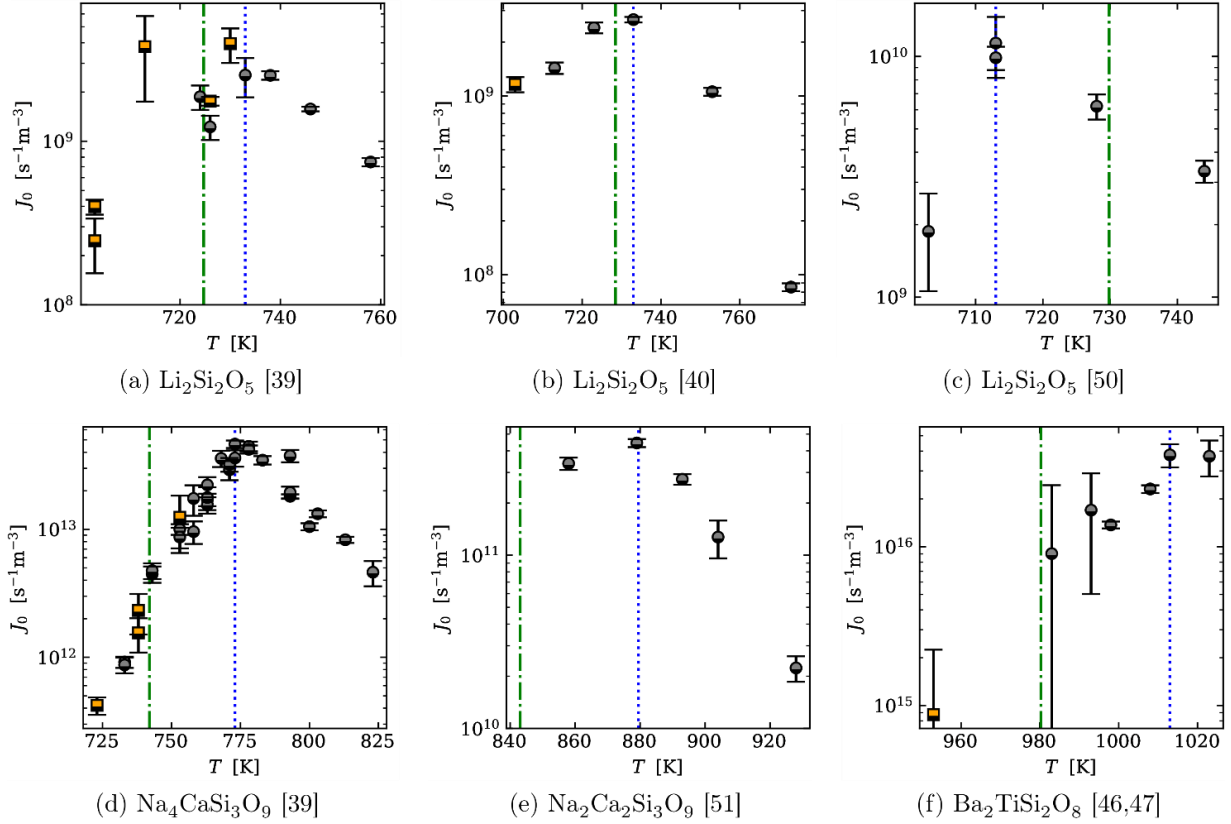

Figure 5  Steady-state crystal nucleation rates versus temperature obtained by fitting $N_V$ data. Datasets that passed the steady-state test are shown as gray circles, whereas datasets that did not pass the test are shown as orange squares. The uncertainty is two standard deviations (confidence of about 95%). The vertical dotted lines show $T_{max}$, whereas the dash-dotted lines show $T_{12}$.

## 5.2 Searching for evidence of the nucleation "break"

Figure 6 and Figure 7 show the results of the numerical calculations performed to check for the nucleation "break". The first assumes that σ is temperature-independent, whereas the latter does not rely on such an assumption. A sign of the "break" would be a deviation from linear behavior in Figure 6, or a non-monotonic increase of σ with respect to the temperature in Figure 7.

Considering only the data that passed the steady-state test, there is no sign of the nucleation "break" for all but one subplot in Figure 6 (subplot d) and one in Figure 7 (subplot c), within the uncertainty margin and confidence bands. Even if we consider the data that have not passed the steady-state test, then the only difference is that Figure 7a also shows a sign of the nucleation "break". We will discuss these results in the following section.

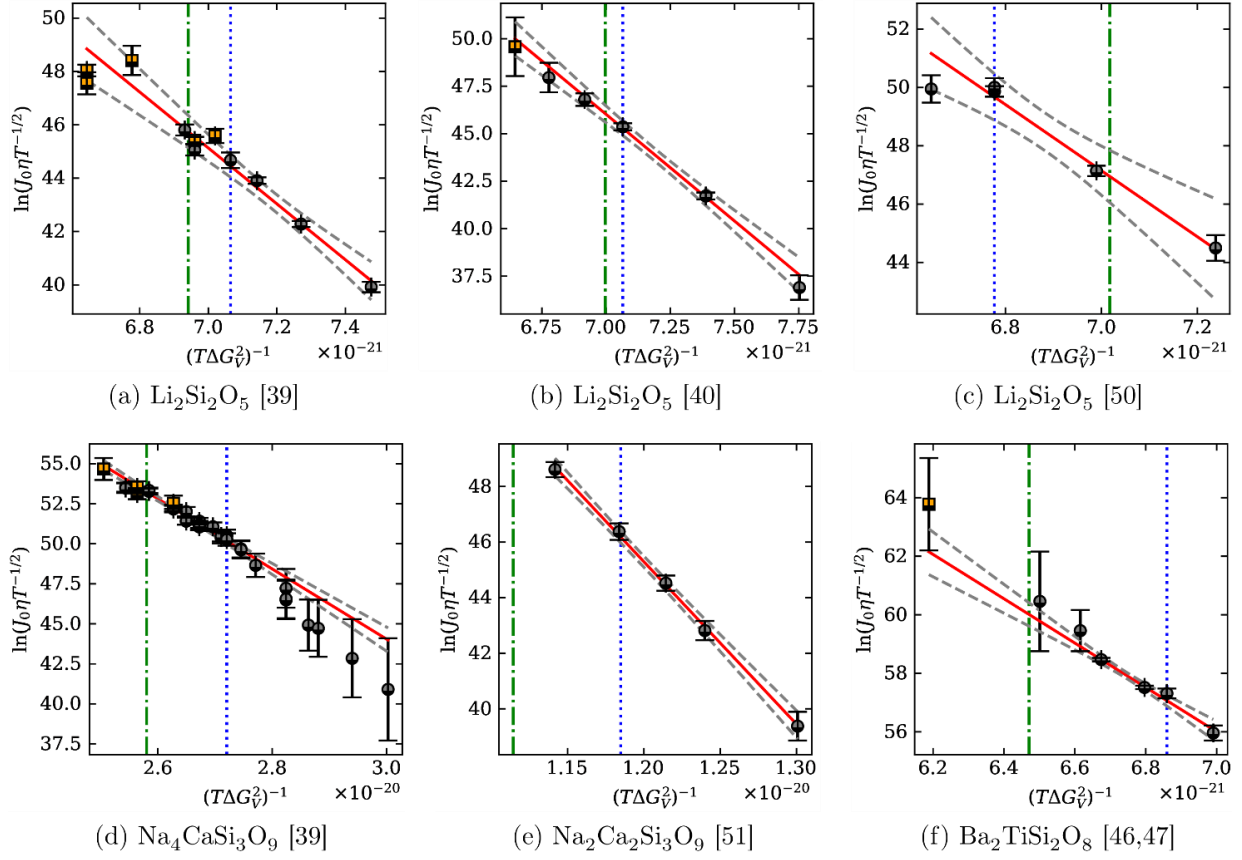

Figure 6  Analysis of nucleation rate data for different materials considering $D_J = D_\eta$ and assuming that $\sigma$ is temperature-independent. Datasets that passed the steady-state test are shown as gray circles, whereas datasets that did not pass the test are shown as orange squares. The continuous red line is the linear regression of the data that passed the steady-state test, and the confidence bands are given by the dashed gray line. The vertical dotted lines mark $T_{max}$, and the dash-dotted lines mark $T_{12}$. The uncertainty of the data is two standard deviations (confidence of 95%).

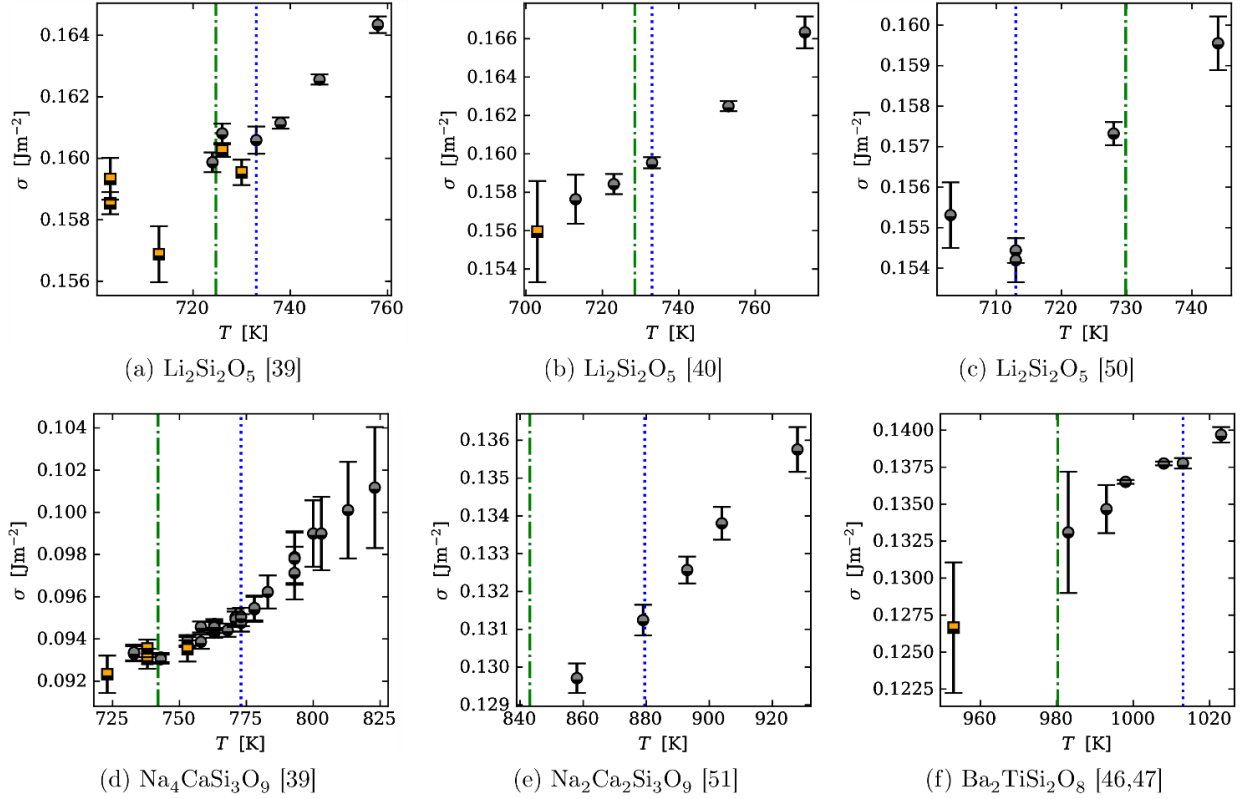

Figure 7    Analysis of nucleation rate data for different materials considering $D_J = D_\eta$ and assuming that σ is temperature-dependent. Datasets that passed the steady-state test are shown as gray circles, whereas datasets that did not pass the test are shown as orange squares. The vertical dotted lines mark $T_{max}$, and the dash-dotted lines mark $T_{12}$. The uncertainty of the data is two standard deviations (confidence of 95%).

# 6 Discussion

Let us begin by analyzing Figure 6. We studied three "clean" datasets for lithium disilicate ($Li_2Si_2O_5$), which are shown in subplots 6a to 6c. Subplot 6b is perhaps the easiest to analyze in this set: there is simply no sign of the nucleation "break" whatsoever, with very small scatter of the data, and only one data point that did not pass the steady-state test. Even this last data point is well within the confidence bands of the linear regression. Just to remind the reader, the non-steady-state points were not used for the linear regressions shown in Figure 6.

The data in subplot 6a is visually more scattered, with a total of five data points that did not pass the steady-state test. Even with this wide scatter, all data points are within the confidence bands of the regression, so we cannot reject that they have a linear dependency within the plot scale. Even the non-steady-state data are within the confidence bands. Once more there is no sign of the nucleation "break".

Subplot 6c, however, is the first result for which we cannot completely reject the presence of the nucleation "break". The data point for the lowest temperature lies in the border of the confidence bands, perhaps not following the expected linear trend. However, two pieces of information are important to mention: *i*) the temperature of the "break" in this plot is significantly below the temperature previously reported for lithium disilicate [10–14]; *ii*) the dispersion of the $N_V$ versus $t$ data for the lowest temperature point is significant (see supplementary material), much higher than for any other $N_V$ dataset.

Similar to subplot 6c, subplot 6d shows another result (but for a different composition, $Na_4CaSi_3O_9$) for which we cannot reject the nucleation "break". The massive amount of data points for this particular glass indicates that there might be a "break" at $T_{max}$, with two different linear behaviors above and below this temperature. Subplots 6e for $Na_2Ca_2Si_3O_9$ and 6f for $Ba_2TiSi_2O_8$ show no sign of the nucleation "break". The results obtained by analyzing Figure 6 do not change if we also consider the data points that did not pass the steady-state test.

The previous paragraphs focused on the results shown in Figure 6, which is an analysis that assumes that the interfacial energy between the critical nuclei and the ambient phase is temperature-independent. Differently, Figure 7 shows the results obtained without making such an assumption. In the subplots of Figure 7, we seek evidence for the nucleation "break" in the form of a change in the monotonic growth of σ regarding the temperature. If we consider only the data points that passed the steady-state test, we observe that only subplot 7c may have a non-monotonic growth of σ. Even subplot 7d, for composition $Na_4CaSi_3O_9$, has no sign of the "break" in this analysis. This is a conflicting result when we compare Figures 6d and 7d, as the "break" was clearer in Figure 6d. When considering the data points that have not passed the steady-state test, then subplot 7a also shows a sign of the "break".

While we believe that it is relevant to test if the steady-state regime was achieved, our results were only slightly affected by limiting the analysis to steady-state data. Therefore, we have not found a strong basis to recommend such practice. We attribute the contrast between our results (very weak evidence for the "break") and those published in the literature (strong evidence for the "break") for the studied compositions [10–14] to the "clean" dataset approach and to the statistical procedure of uncertainty propagation combined with confidence bands used here.

In the end, having discussed the procedures and the current results in this communication, we are not able to draw a definitive conclusion regarding the existence or not of the nucleation "break". Our results, however, cast reasonable doubt about whether the "break" is a phenomenon that happens for all oxide glasses. A recent report by Xia et al. [59], with new measurements of the nuclei density in a $Ba_5Si_8O_{21}$ glass at 50 K below $T_{max}$, supports our conclusion that the nucleation "break" is an artifact.

For the cases where the "break" was not evident, one could argue that perhaps it happens in a temperature even below the temperature range for which nucleation data are available. This hypothesis, however, can only be tested by collecting new kinetic data at even lower temperatures for very long times—an endeavor that we are currently working on for four glass-forming oxide systems.

A common conjecture by experts in this field is that some kind of "break" should indeed happen at a temperature when the critical nucleus size becomes equal to one unit cell or the jump distance, $d_0$. A result that follows, considering (for a rough estimate) that σ is temperature-independent, is that in no temperature between 1 K and the glass transition temperature, the critical nucleus size, $R^*$, is equal to the jump distance for the studied compositions. This scenario changes if σ is considered to have a linear, positive temperature dependence. In doing so, by extrapolation, a finite temperature where $R^* = d_0$ arises. We called this temperature $T_\Omega$. The values of $T_\Omega$ are significantly below $T_{12}$ for all compositions studied here, as can be confirmed by the data shown in Table 1. We thus believe this route for explaining a possible nucleation break—that is, that the proximity of $T_\Omega$ with the region where experimental nucleation data is available—is weakened.

Overall, our findings tackle a long-standing problem regarding the Classical Nucleation Theory. We believe that shedding light on this alleged failure of the Classical Nucleation Theory is indeed a critical step toward the understanding and prediction of crystal nucleation kinetics in glass-forming systems. We emphasize the importance of carrying out extensive measurements for long times below $T_{max}$ for other systems following the recommendations of this work, and reporting all the metadata associated with nucleation experiments, specially the thermal history, including the heating and cooling rates of the samples. Such experiments should close this case.

# 7 Summary and Conclusions

We analyzed nucleation rate data ($N_v$ versus time) in depth for six clean datasets of four glass-forming systems, seeking evidence to support or discard the alleged failure of the Classical Nucleation Theory below $T_{max}$.

Our analysis followed a new, rigorous protocol based on three items: *i*) Only the $N_v$ versus time data that passed a steady-state regime test were used; *ii*) Nucleation and viscosity data were measured for samples of the same batch, and each batch was individually analyzed; and *iii*) the uncertainty and regression confidence bands were computed and considered. With this setup, our results cast reasonable doubt on whether the nucleation "break" is a universal phenomenon that happens for all oxide glasses.

### CRediT author statement

Daniel Roberto Cassar: Conceptualization, Methodology, Software, Formal analysis, Data Curation, Writing - Original Draft, Writing - Review & Editing, Visualization. André Hofmeister Serra: Investigation. Oscar Peitl: Investigation. Edgar Dutra Zanotto: Conceptualization, Resources, Writing - Review & Editing, Supervision, Project administration, Funding acquisition.

### Acknowledgments

This study was financed by the Brazilian agencies National Council for Scientific and Technological Development (CNPq), São Paulo State Research Foundation (FAPESP grant numbers 2017/12491-0 and 2013/07793-6), and by the Coordenação de Aperfeiçoamento de Pessoal de Nível Superior - Brasil (CAPES) - Finance Code 001. Discussions with V.M. Fokin and A.S. Abyzov are greatly appreciated. We would also like to thank D. Kashchiev for kindly reaching out with relevant information that we used to review and update our analysis, and G. Sycheva and A. Cabral for kindly providing digital copies of their theses.

# Supplementary material to "Critical assessment of the alleged failure of the Classical Nucleation Theory at low temperatures"

Daniel R. Cassar, André H. Serra, Oscar Peitl, Edgar D. Zanotto

*Department of Materials Engineering, Federal University of* São *Carlos*

*Graduate Program in Materials Science and Engineering, São Carlos, SP, Brazil*

# 1 $N_V$ analysis of $Li_2Si_2O_5$ (Ref. [39])

Figure S1 shows the $N_V$ versus $t$ data measured by Fokin [39] with regression by the Kashchiev expression, Eq. (5). Figure S2 shows the same data, with the addition of the asymptotic steady-state line. The intention of Figure S2 is to show a visual way to check which datasets are close or far from the steady-state.

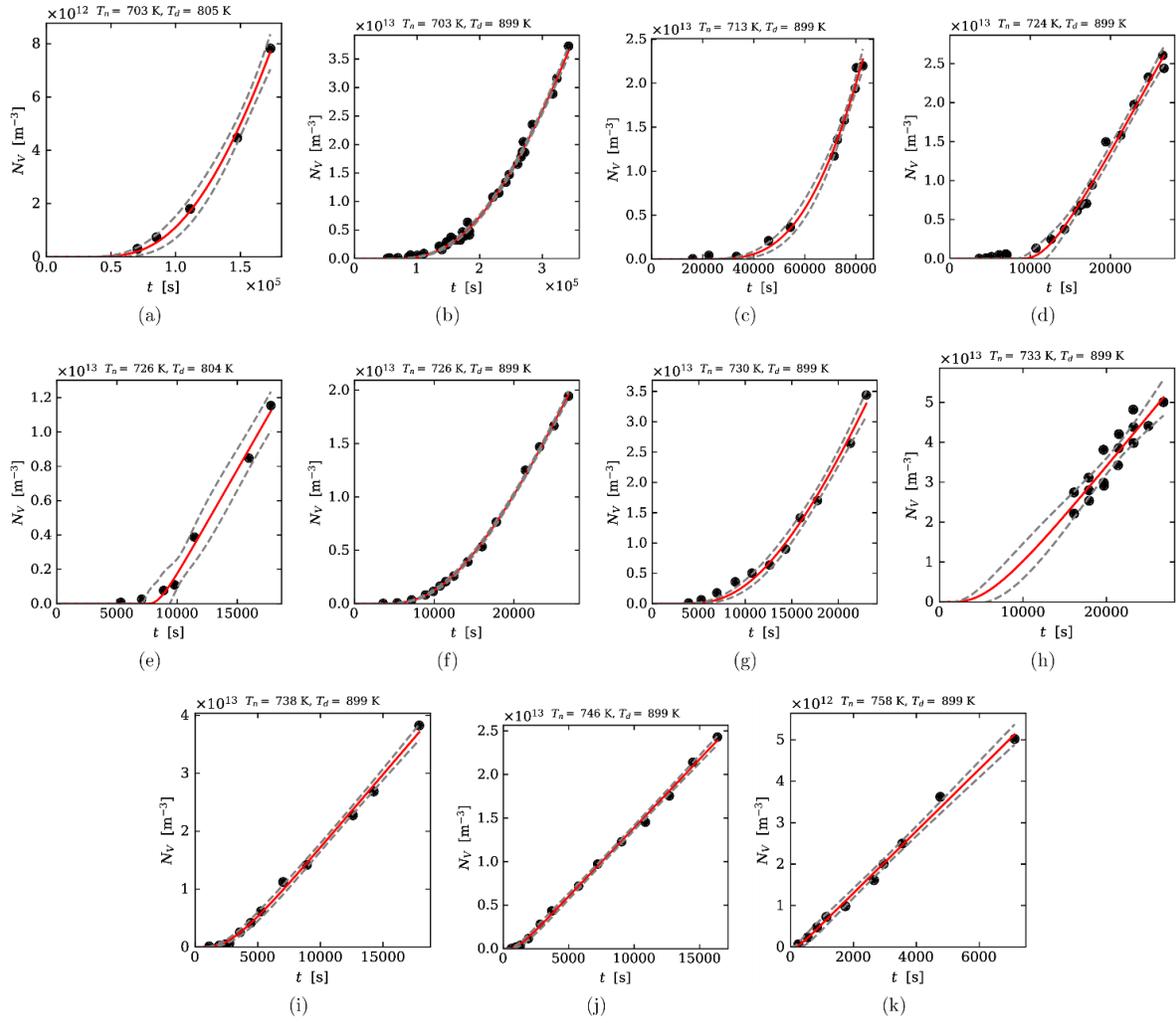

Figure S1   $N_V$ versus $t$ for $Li_2Si_2O_5$ measured in Ref. [39]. The solid red line is the regression of the data using the Kashchiev equation (5) and the confidence bands are shown by a dashed gray line.

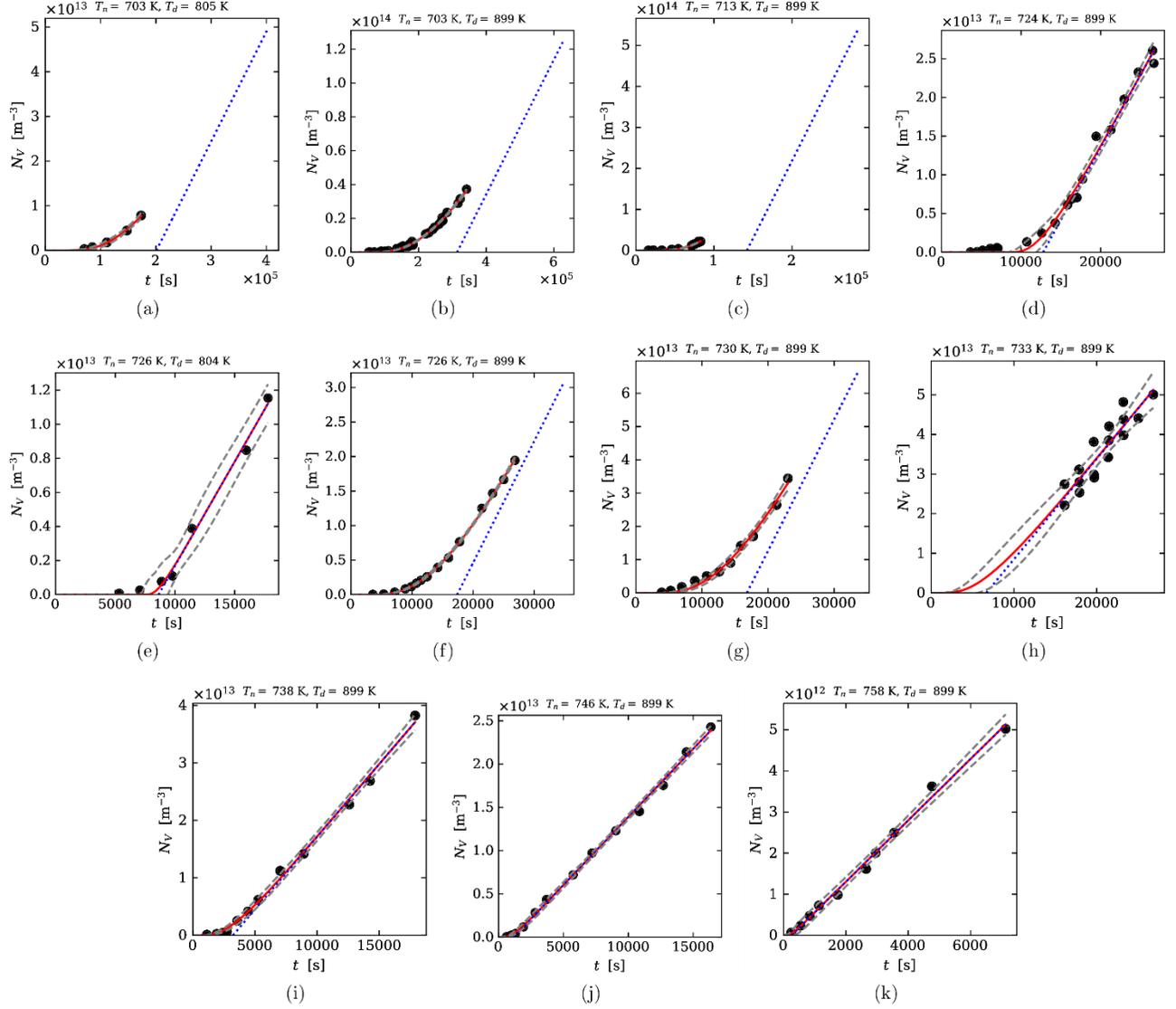

Figure S2   $N_V$ versus $t$ for $Li_2Si_2O_5$ measured in Ref. [39]. The solid red line is the regression of the data using the Kashchiev equation (5) and the confidence bands are shown by a dashed gray line. The dotted blue line is the asymptotic steady-state line with a slope of $J_0$ that intercepts the $x$-axis at $t_{\text{ind,d}}$.

## 2 $N_V$ analysis of $Li_2Si_2O_5$ (Ref. [40])

Figure S3 shows the $N_V$ versus $t$ data measured by Sycheva [40] with the regression of the Kashchiev equation (5). Figure S4 shows the same data, with the addition of the asymptotic steady-state line.

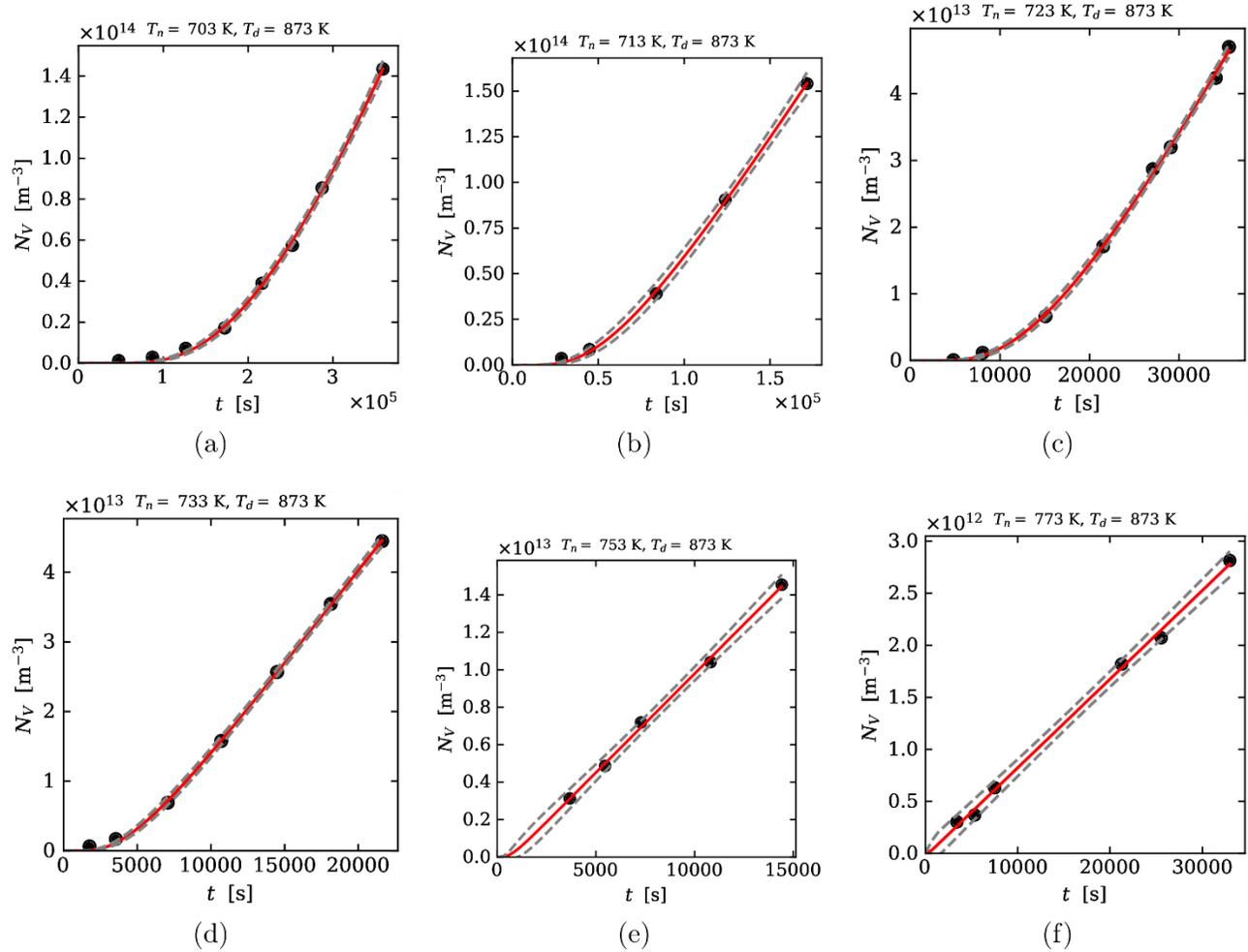

Figure S3  $N_V$ versus $t$ for $Li_2Si_2O_5$ measured in Ref. [40]. The solid red line is the regression of the data using the Kashchiev equation (5) and the confidence bands are shown by a dashed gray line.

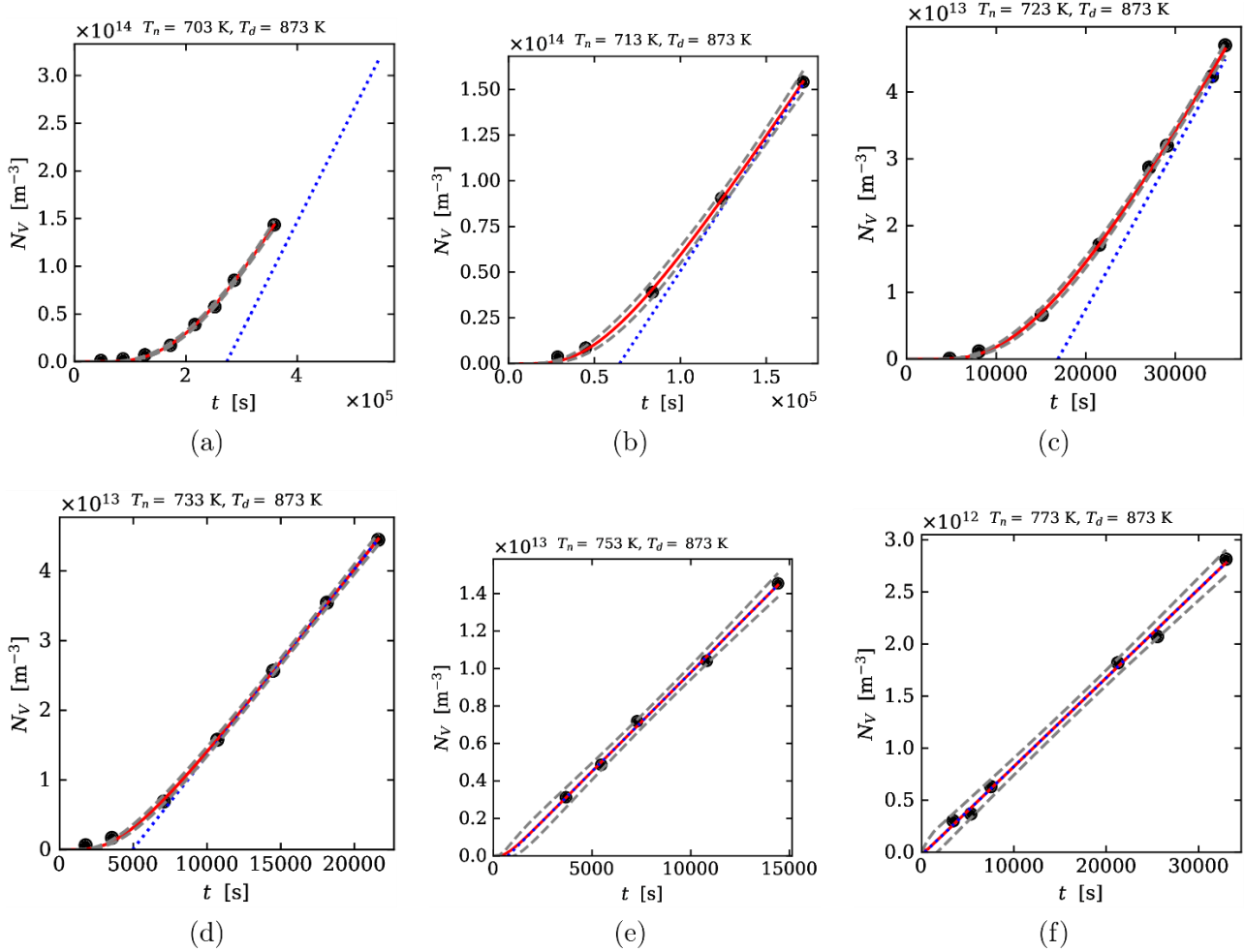

Figure S4  $N_V$ versus $t$ for $Li_2Si_2O_5$ measured in Ref. [40]. The solid red line is the regression of the data using the Kashchiev equation (5) and the confidence bands are shown by a dashed gray line. The dotted blue line is the asymptotic steady-state line with a slope of $J_0$ that intercepts the $x$-axis at $t_{ind,d}$.

# 3  $N_V$ analysis of Li$_2$Si$_2$O$_5$ (Ref. [50])

Figure S5 shows the $N_V$ versus $t$ data measured by Serra [50] with the regression of the Kashchiev equation (5). Figure S6 shows the same data, with the addition of the asymptotic steady-state line.

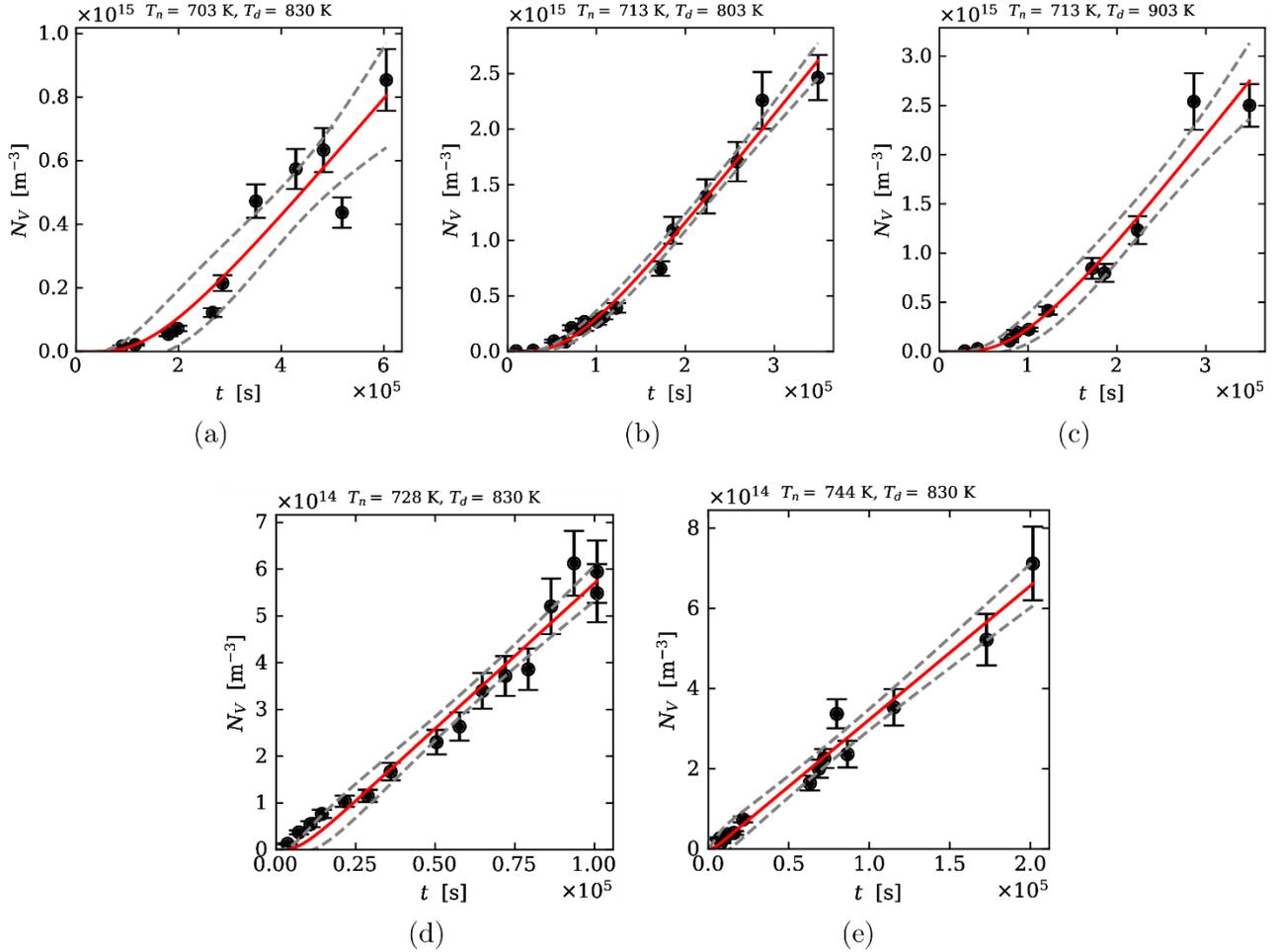

Figure S5  $N_V$ versus $t$ for Li$_2$Si$_2$O$_5$ measured in Ref. [50]. The solid red line is the regression of the data using the Kashchiev equation (5) and the confidence bands are shown by a dashed gray line.

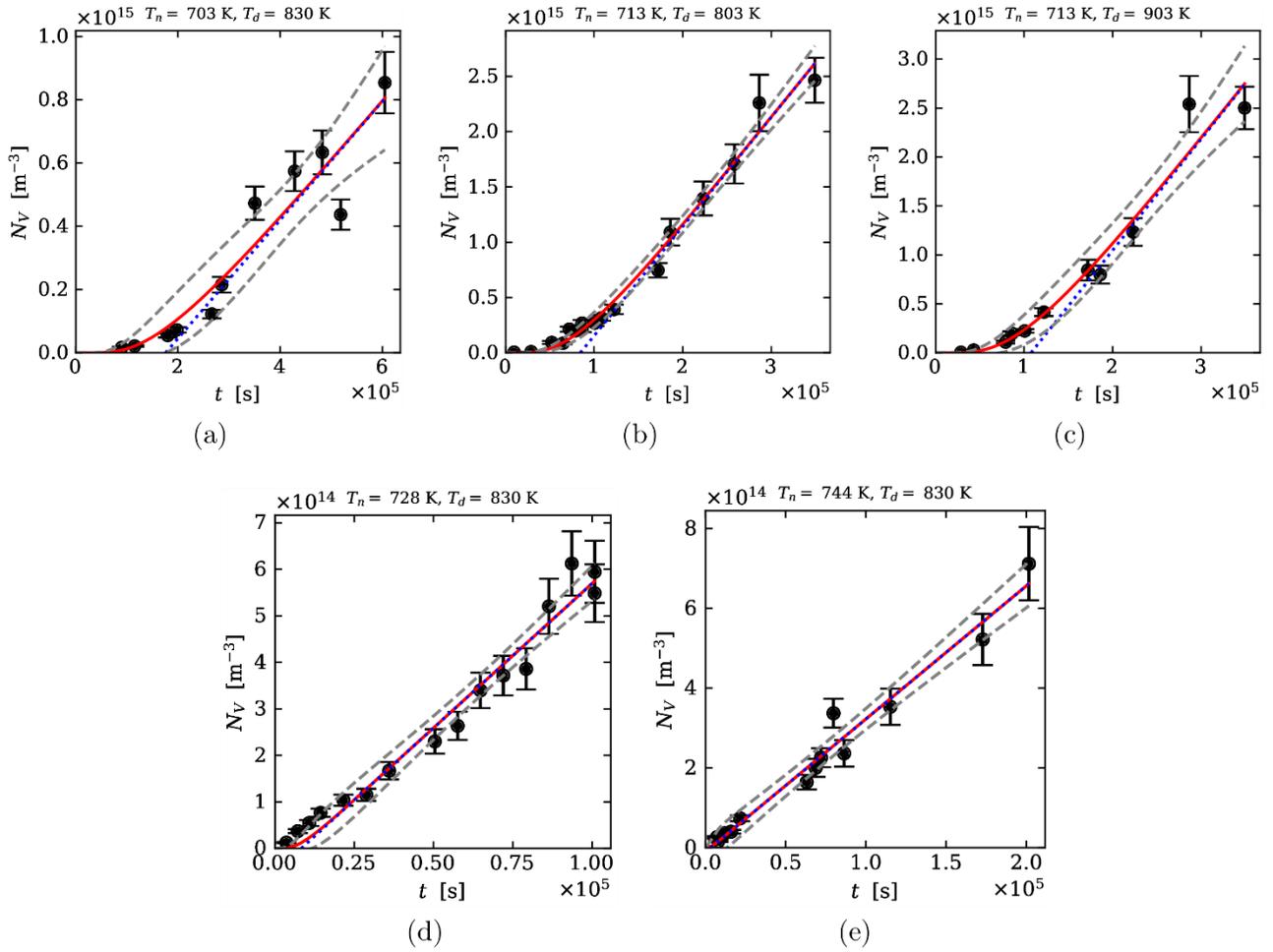

Figure S6  $N_V$ versus $t$ for $Li_2Si_2O_5$ measured in Ref. [50]. The solid red line is the regression of the data using the Kashchiev equation (5) and the confidence bands are shown by a dashed gray line. The dotted blue line is the asymptotic steady-state line with a slope of $J_0$ that intercepts the $x$-axis at $t_{\text{ind,d}}$.

# 4  $N_V$ analysis of Na$_4$CaSi$_3$O$_9$ (Ref. [39])

Figure S7 and Figure S8 show the $N_V$ versus $t$ data measured by Fokin [39] with the regression of the Kashchiev equation (5). Figure S9 and Figure S10 show the same data, with the addition of the asymptotic steady-state line.

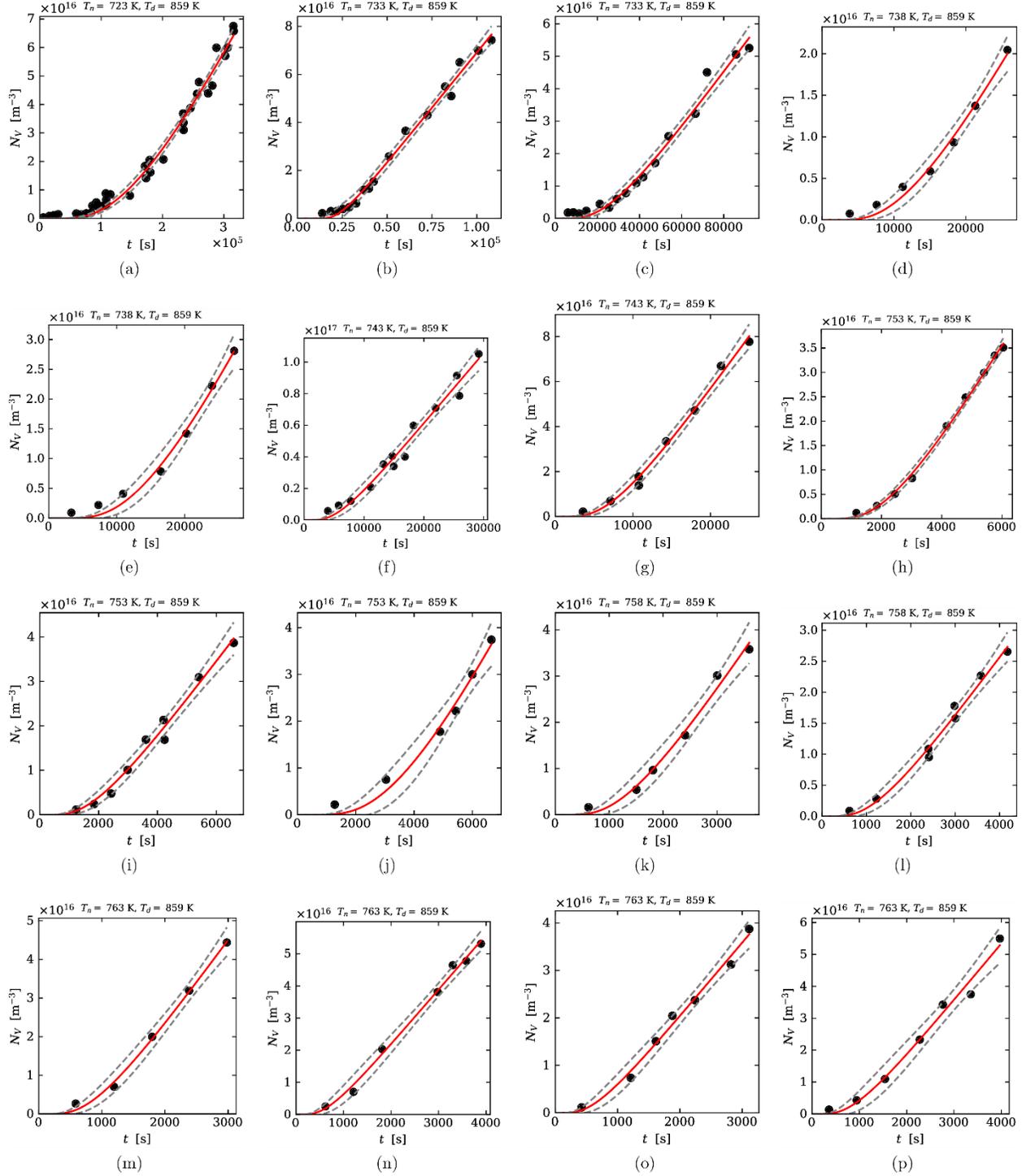

Figure S7  $N_V$ versus $t$ for Na$_4$CaSi$_3$O$_9$ measured in Ref. [39] for temperatures below 763 K. The solid red line is the regression of the data using the Kashchiev equation (5) and the confidence bands are shown by a dashed gray line.

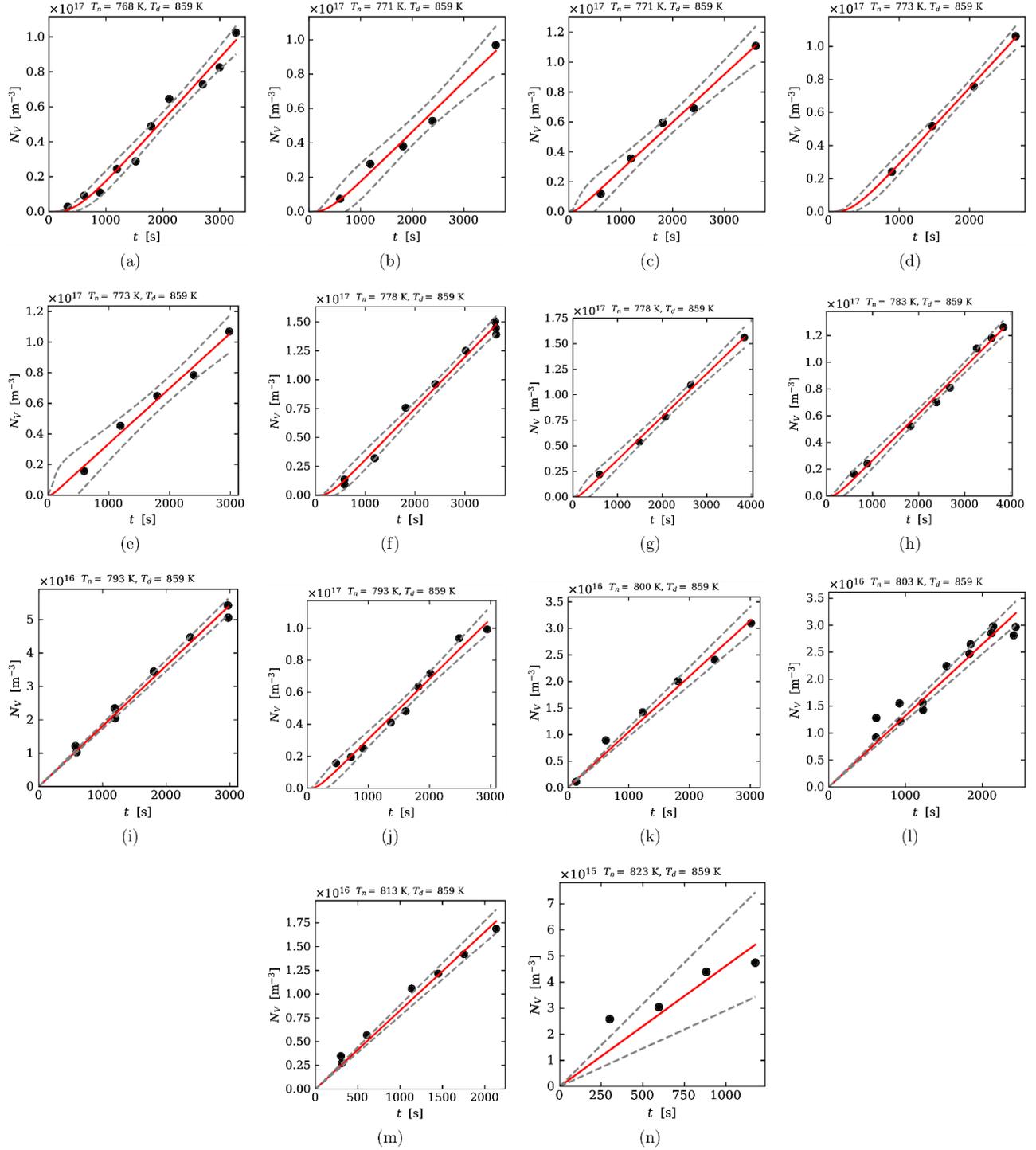

Figure S8  $N_V$ versus $t$ for $Na_4CaSi_3O_9$ measured in Ref. [39] for temperatures above 768 K. The solid red line is the regression of the data using the Kashchiev equation (5) and the confidence bands are shown by a dashed gray line.

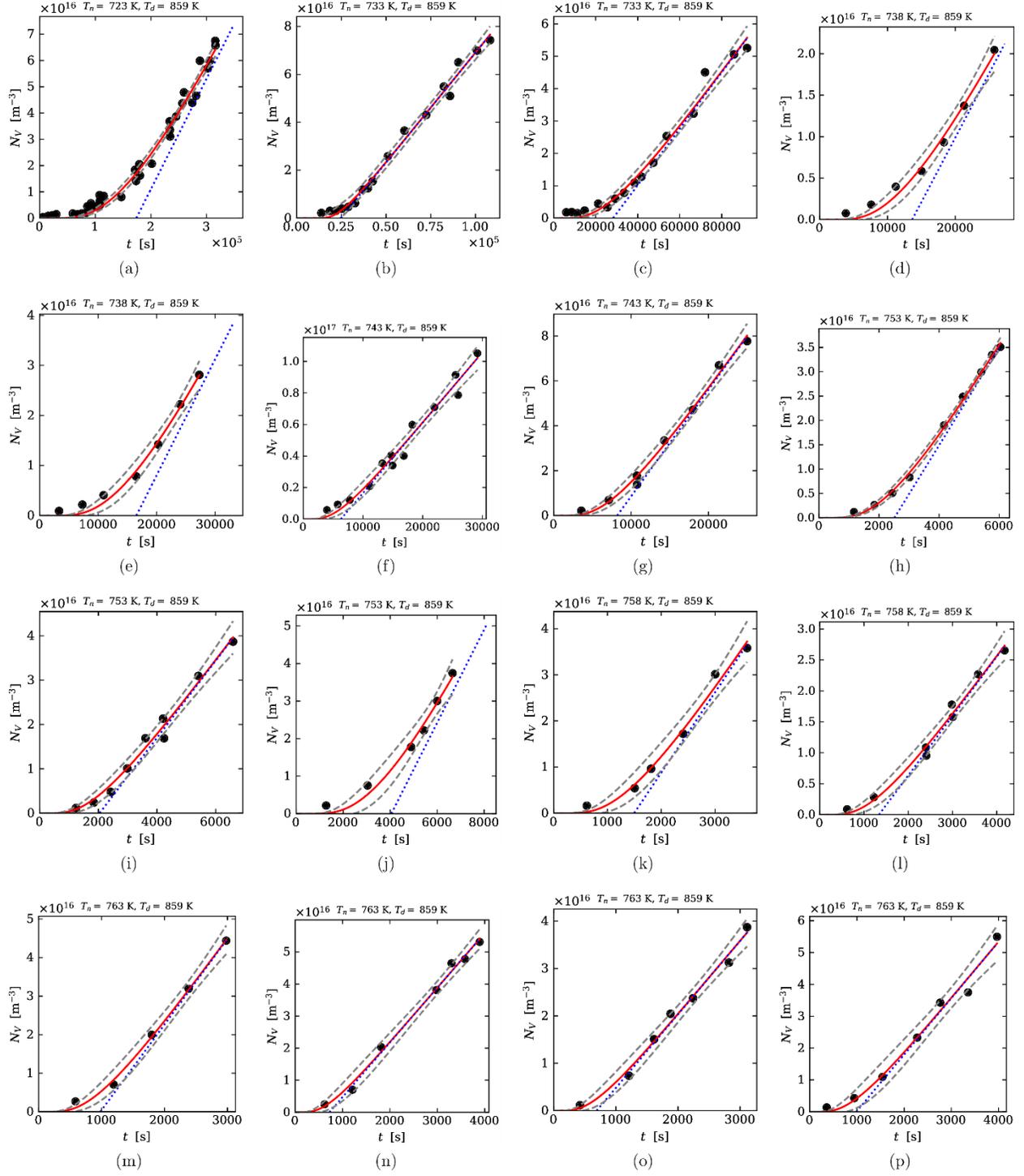

Figure S9   $N_V$ versus $t$ for $Na_4CaSi_3O_9$ measured in Ref. [39] for temperatures below 763 K. The solid red line is the regression of the data using the Kashchiev equation (5) and the confidence bands are shown by a dashed gray line. The dotted blue line is the asymptotic steady-state line with a slope of $J_0$ that intercepts the $x$-axis at $t_{\text{ind,d}}$.

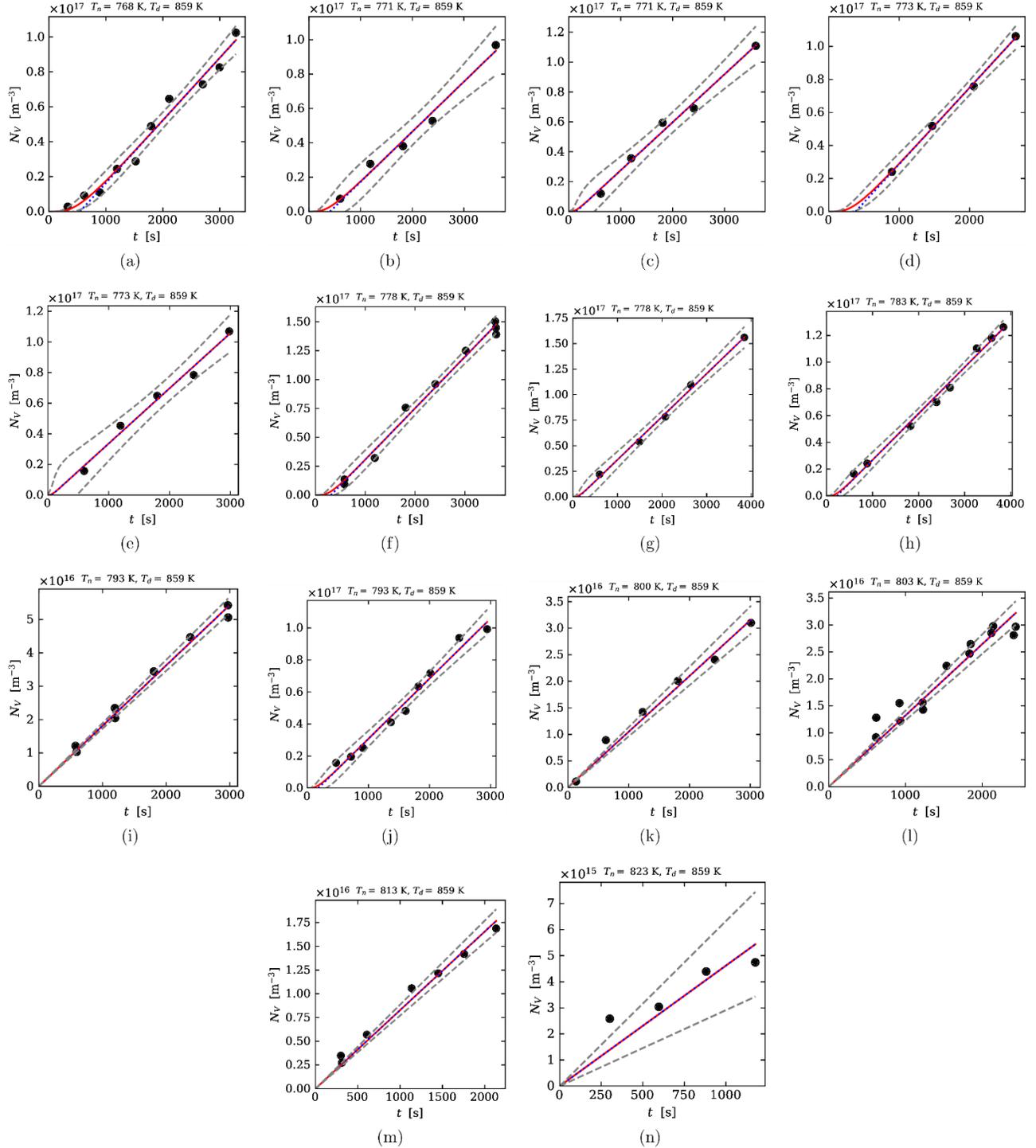

Figure S10 $N_V$ versus $t$ for $Na_4CaSi_3O_9$ measured in Ref. [39] for temperatures above 768 K. The solid red line is the regression of the data using the Kashchiev equation (5) and the confidence bands are shown by a dashed gray line. The dotted blue line is the asymptotic steady-state line with a slope of $J_0$ that intercepts the $x$-axis at $t_{ind,d}$.

# 5 $N_V$ analysis of Na$_2$Ca$_2$Si$_3$O$_9$ (Ref. [51])

Figure S11 shows the $N_V$ versus $t$ data measured by Gonzalez-Oliver [51] with the regression of the Kashchiev equation (5). Figure S12 shows the same data, with the addition of the asymptotic steady-state line.

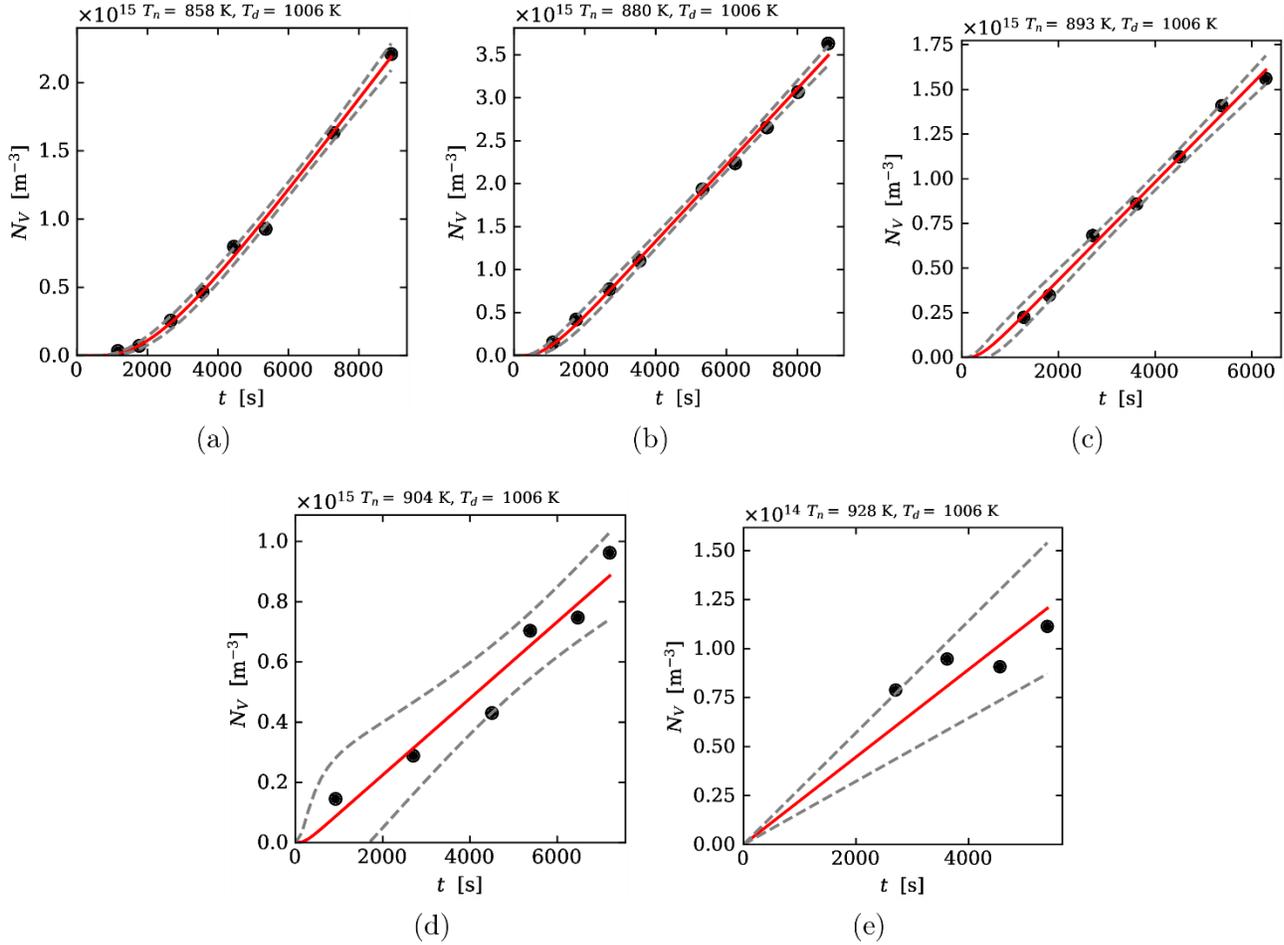

Figure S11 $N_V$ versus $t$ for Na$_2$Ca$_2$Si$_3$O$_9$ measured in Ref. [51]. The solid red line is the regression of the data using the Kashchiev equation (5) and the confidence bands are shown by a dashed gray line.

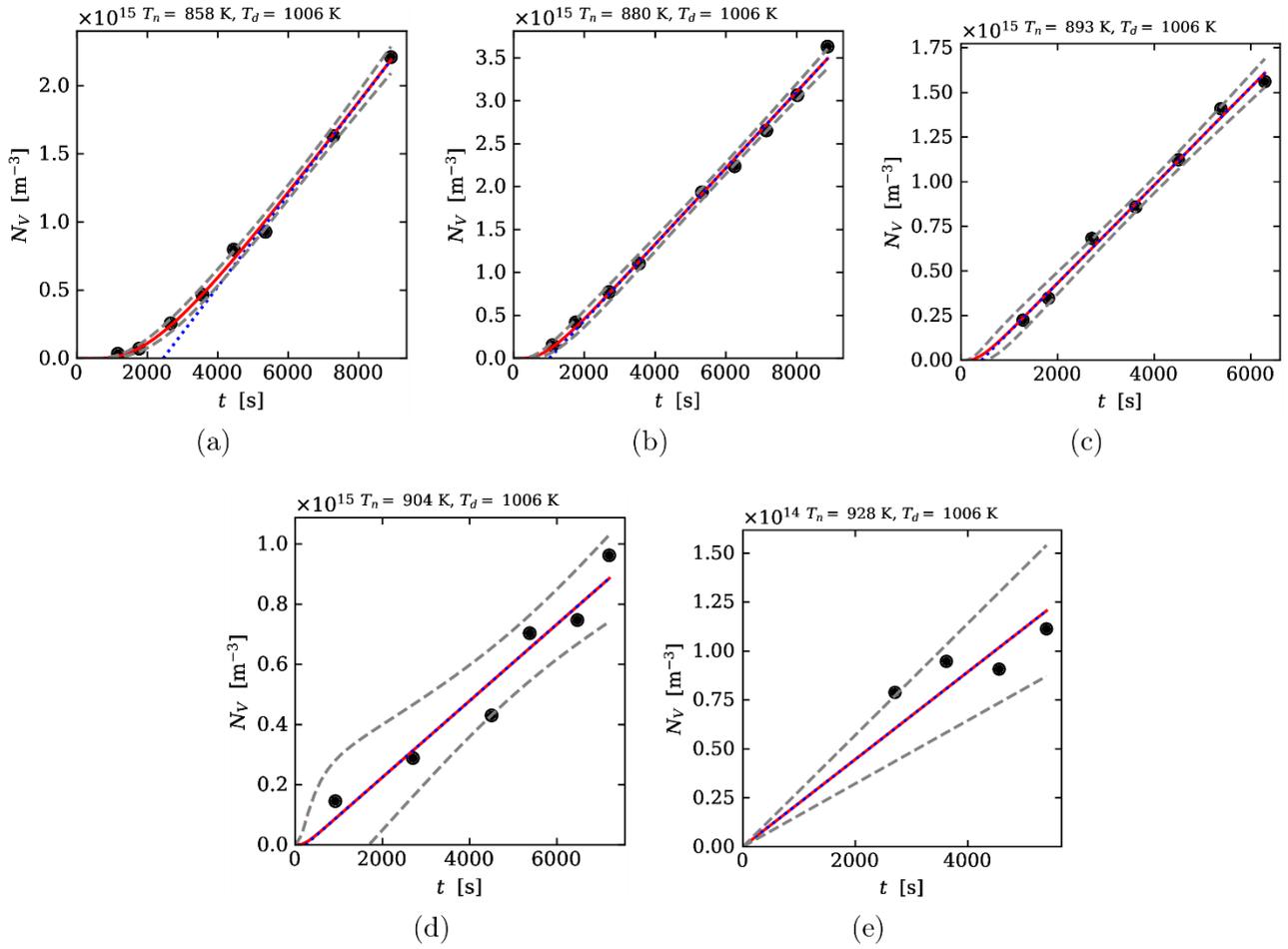

Figure S12 $N_V$ versus $t$ for $Na_2Ca_2Si_3O_9$ measured in Ref. [51]. The solid red line is the regression of the data using the Kashchiev equation (5) and the confidence bands are shown by a dashed gray line. The dotted blue line is the asymptotic steady-state line with a slope of $J_0$ that intercepts the $x$-axis at $t_{ind,d}$.

# 6 $N_V$ analysis of $Ba_2TiSi_2O_8$ (Ref. [46])

Figure S13 shows the $N_V$ versus $t$ data measured by Cabral [46] with the regression of the Kashchiev equation (4). Figure S14 shows the same data, with the addition of the asymptotic steady-state line.

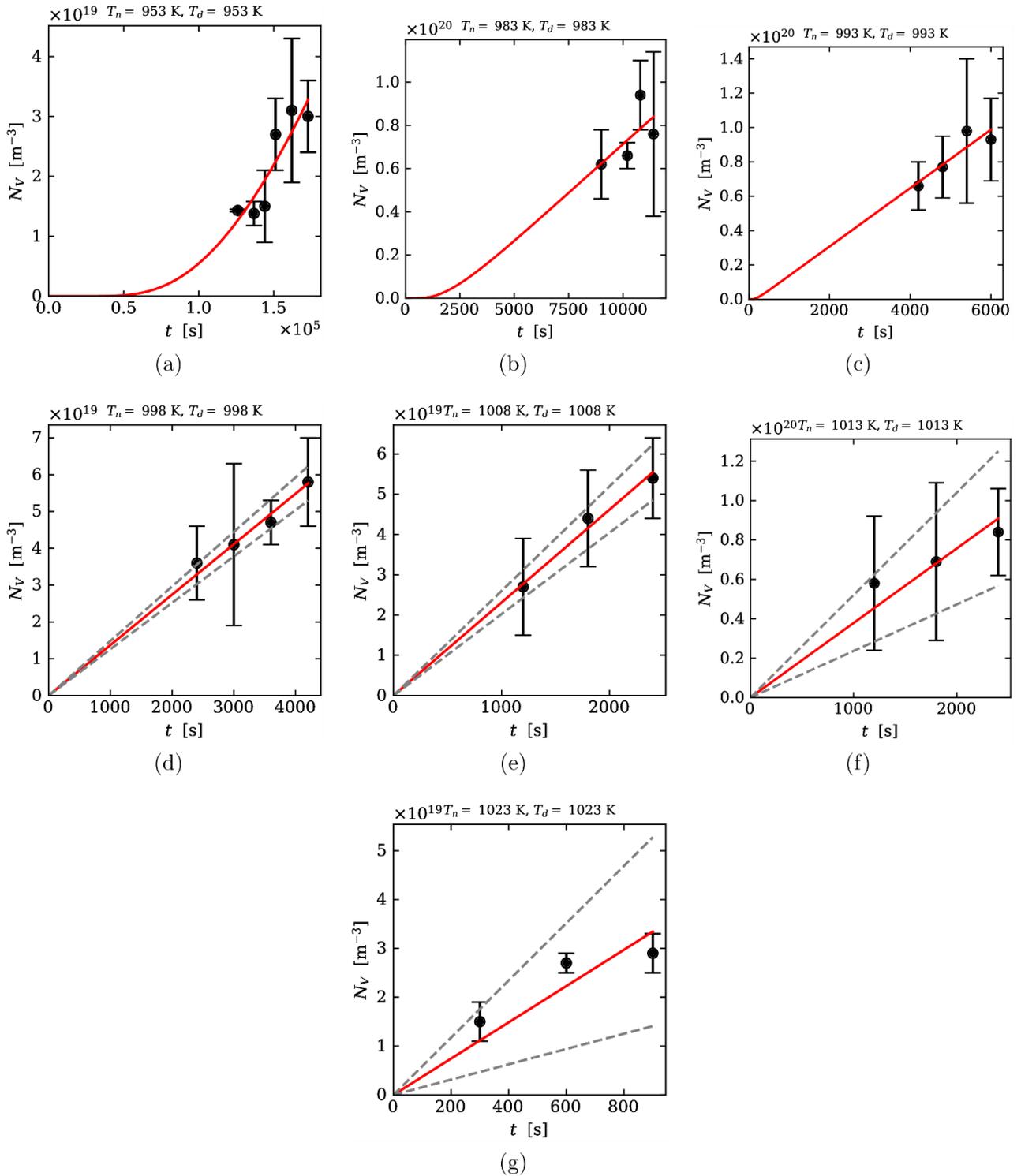

Figure S13 $N_V$ versus $t$ for $Ba_2TiSi_2O_8$ measured in Ref. [46]. The solid red line is the regression of the data using the Kashchiev equation (4) and the confidence bands are shown by a dashed gray line.

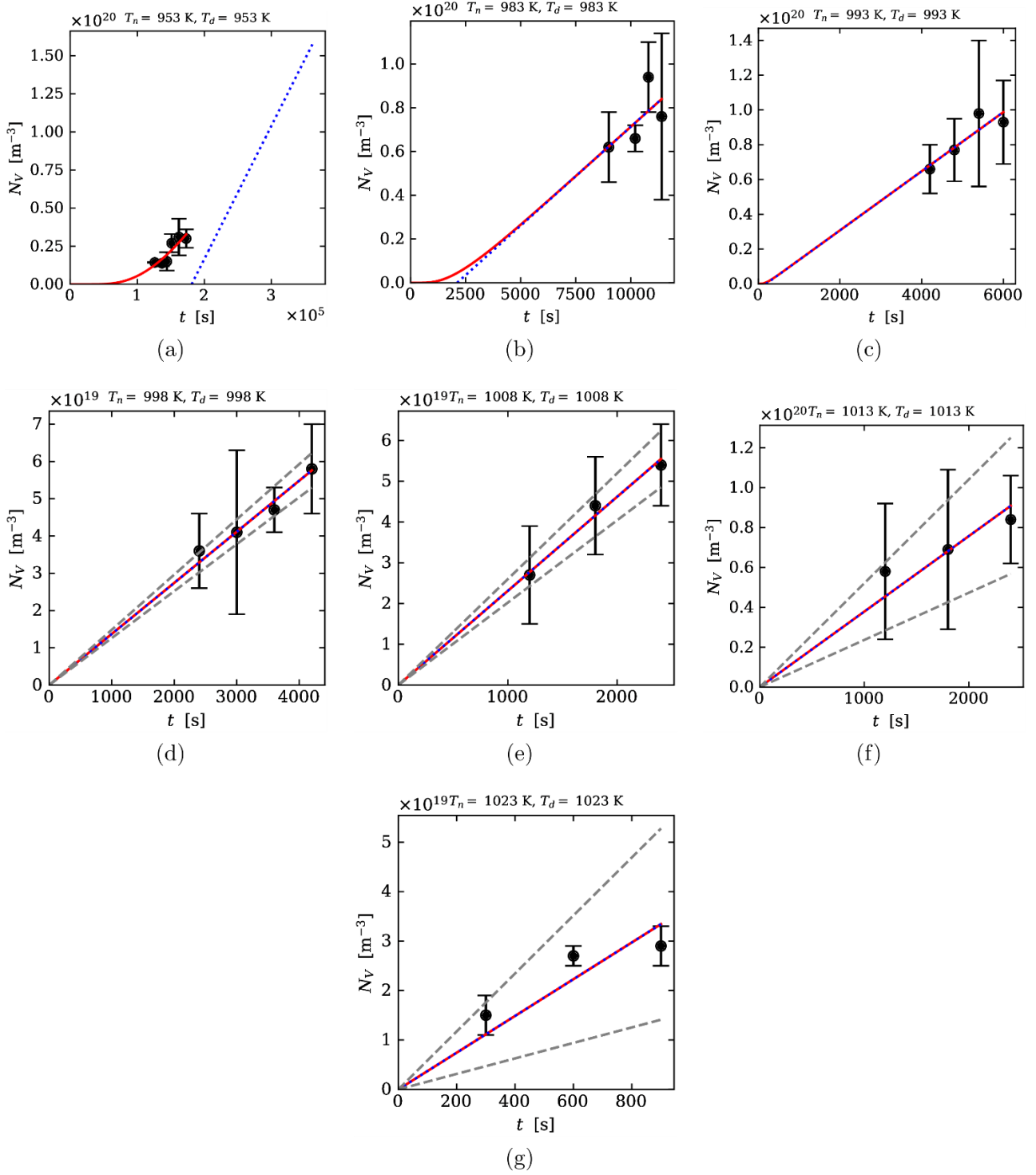

Figure S14 $N_V$ versus $t$ for $Ba_2TiSi_2O_8$ measured in Ref. [46]. The solid red line is the regression of the data using the Kashchiev equation (4) and the confidence bands are shown by a dashed gray line. The dotted blue line is the asymptotic steady-state line with a slope of $J_0$ that intercepts the $x$-axis at $t_{ind,n}$.